\documentclass{article}

\usepackage{geometry}
\usepackage{cite}
\usepackage{amsmath,amssymb,amsfonts}

\usepackage{subfigure}
\usepackage{graphicx}
\usepackage{mathtools}
\usepackage{xcolor}
\usepackage{amsfonts}
\usepackage{amssymb}
\usepackage{adjustbox}
\usepackage{multirow}
\newcommand{\kmer}[0]{$k$-mer}
\newcommand{\kmers}[0]{$k$-mers}
\newcommand{\ours}{ViQUF}
\def\BibTeX{{\rm B\kern-.05em{\sc i\kern-.025em b}\kern-.08em
    T\kern-.1667em\lower.7ex\hbox{E}\kern-.125emX}}

\title{ViQUF: de novo Viral Quasispecies reconstruction using Unitig-based Flow
networks}

\author{Borja~Freire$^1$, 
        Susana~Ladra$^1$, 
        Jos{\'e}~R.~Param{\'a}$^1$,  
         Leena Salmela$^2$\\
\small $^1$ Universidade da Coru{\~n}a, Centro de investigaci{\'on} CITIC, Facultad de Inform{\'a}tica, 15071,\\ \small A Coru{\~n}a, Spain.
\\
\small E-mail: \{borja.freire1,susana.ladra,jose.parama\}@udc.es\\
\small $^2$Department of Computer Science, Helsinki Institute for Information Technology, \\ \small University of Helsinki, Helsinki, Finland \protect\\
\small E-mail: leena.salmela@cs.helsinki.fi} 

\date{}

\usepackage{fancyhdr}
\begin{document}

\maketitle

\abstract{ During viral infection, intrahost mutation and recombination can lead to significant evolution, resulting in a population of viruses that harbor multiple haplotypes. The task of reconstructing these haplotypes from short-read sequencing data is called viral quasispecies assembly, and it can be categorized as a multiassembly problem. 
We consider the  \textit{de novo}  version of the problem, where no reference is available.
 We present ViQUF, a \textit{de novo} viral quasispecies assembler that addresses haplotype assembly and quantification. ViQUF obtains a first draft of the  assembly graph from a  de Bruijn graph. Then, solving a min-cost flow over a flow network built for each pair of adjacent vertices based on their paired-end information creates an \textit{approximate paired assembly  graph} with suggested frequency values as edge labels, which is the first frequency estimation. Then, original haplotypes are obtained through a greedy path reconstruction guided by a min-cost flow solution in the \textit{approximate paired assembly graph}. ViQUF outputs the contigs with their frequency estimations. Results on real and simulated data show that ViQUF is at least four times faster using at most half of the memory than previous methods,
while maintaining, and in some cases outperforming, the high quality of assembly and frequency estimation of overlap graph-based methodologies, which are known to be more accurate but slower than the de Bruijn graph-based approaches.  \\
\textbf{Availability:} ViQUF is freely available at: {{https://github.com/borjaf696/ViQUF}}\\

}

\section{Introduction}
A
 host organism infected by an RNA virus, such as human immunodeficiency virus (HIV-1), typically carries a population of related but distinct viral haplotypes, i.e., viral quasispecies \cite{quasispecies,quasispecies2}, due to their high mutation rate \cite{mutationrate}. However, it is possible to use short-read sequencing data to reconstruct the set of viral haplotypes in a sample with their relative abundance. In this work, we focus on {\em de novo} viral quasispecies assembly, which does not require a reference sequence.
 There are reference-based methods for viral quasispecies reconstruction \cite{shorah,predicthaplo,cliquesnv}; however, this approach is constrained to well-known viruses \cite{10.1007/978-3-642-39159-0_24} or can be biased for certain studies \cite{savage,chen2021reference}.


{\em De novo} viral quasispecies assembly from shotgun reads   has been tackled either using overlap graphs \cite{savage,pehaplo} or de Bruijn graphs (DBGs) \cite{mlehaplo,viaDBG}. Overlap graph-based approaches have traditionally been considered accurate but require many resources, whereas DBG-based approaches are usually efficient but less accurate. Recently, we  showed that when properly taking advantage of paired-end reads using a paired DBG \cite{pairedDBG}, a DBG-based approach can be as accurate as an overlap graph-based method \cite{viaDBG}.

Network flows have been used to solve many assembly problems in computational biology \cite{euler,stringgraph,traph,10.1007/978-3-540-79450-9_15,skums2013reconstruction,virusvg}. These approaches first construct an assembly graph (e.g., an overlap graph, DBG, or string graph), where each vertex and edge is assigned a coverage based on the number of reads mapping to it. Then, a network flow that explains the coverage of each vertex and edge is found. Finally, the flow is split into paths, which are the contigs reported by the methods. Pevzner \textit{et al.} \cite{euler} and Myers \cite{stringgraph} first suggested using network flows to solve genome assembly. Later, Westbrooks \textit{et al.} \cite{10.1007/978-3-540-79450-9_15} used   network flows in the quasispecies assembly problem. Recently, several other works have also used network flows to address the same problem  \cite{skums2013reconstruction,mancuso,astrovskaya2014inferring,virusvg}.  



In this work, we present \ours, an efficient method for assembly and abundance quantification of viral quasispecies. \ours\ builds an assembly graph using the DBG paradigm and uses network flows to formulate the problem of using paired-end reads to split the assembly graph into an approximate paired assembly graph. We obtain the contigs and their frequencies from the approximate paired assembly graph by solving a network flow problem. Furthermore, we present a mathematically rigorous formulation based on kernel density estimation to determine the threshold for solid \kmer{} abundance when building the DBG. The experimental results show that \ours\ is faster and more memory efficient than previous methods while maintaining the good quality of the resulting assembly and frequency estimation.

\section{Preliminaries}

Given an individual  carrying  a virus with reference genome $G$ and a  set of haplotypes $H=\{h_1, h_2, \ldots, h_{n_h}\}$, each one with a relative frequency $f(h_i)$ (or abundance),
a shotgun sequencing process produces a set of reads $D=\{r_1, \ldots, r_{n_r}\}$, each read of length $l$.


 The genome $G$ is a sequence of base pairs $G=\{bp_1, \ldots bp_n\}$.
Each $h_i\in H$ is $G$ but with three types of changes with respect to the reference genome:  substitutions, insertions, and deletions.




 Reads are of \textit{paired-end} type; thus, each read $r_j\in D$ has two parts: $L(r_j)$ is the left-hand read and $R(r_j)$ is the right-hand read. 
 $L(r_j)$  covers the  base pairs $bp_b,\ldots, bp_e$ and $R(r_j)$ covers the  base pairs $bp_{b'},\ldots, bp_{e'}$ 
 of a haplotype $h_t\in H$.
 The average number of base pairs between $bp_b$ and $bp_e'$ is called the \textit{insert size}.

 Due to sequencing errors, reads might include false changes of base pairs, insertions, and deletions. 
  Given a base pair $bp_s$ of the reference genome $G$, there are on average $C$ reads covering  it; this value is the \textit{coverage} of the sequencing.

From the set of reads $D$, our goal is to reconstruct the haplotypes in $H$ as complete as possible and  estimate  the relative frequency of each haplotype in the sample. 

Our method is based on a DBG $G=(V,E)$. The vertices $V = \{v_1,...,v_n\}$ are $k$-mers, i.e., all subsequences of length $k$ of the reads in $D$, and there is an edge $e=(v_i,v_j)$ between two vertices $v_i$ and $v_j$  if the last $k-1$ bases of $v_i$ match the first $k-1$ bases of $v_j$. $k$ is a parameter of the assembly.

\section{Our method: \ours}

\begin{figure*}[t]
	\centering
	\includegraphics[width=0.9\textwidth]{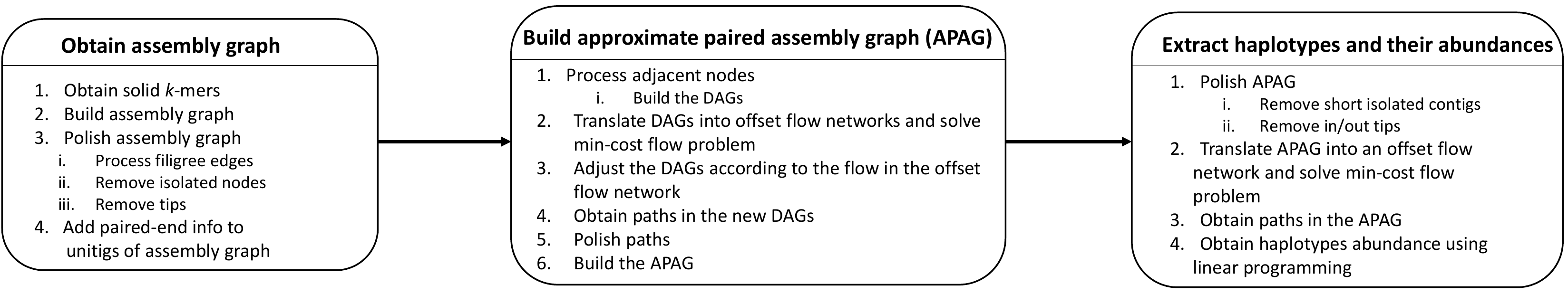}
	\caption{Overview of the proposed method process. }\label{fig:overall}
\end{figure*}

Figure \ref{fig:overall} shows an overview of the steps of ViQUF. The method selects  the  \kmers\ found in the reads in $D$ whose frequency is above a threshold to build a DBG. In that graph, the nonbranching paths are compacted into single nodes, unitigs,  producing an assembly graph (AG). For each unitig, we associate a set of unitigs linked to it by paired-end reads.
Then, AG is processed by taking every pair of adjacent nodes. For each pair, a new directed acyclic graph (DAG) is built including all unitigs linked to those two nodes by paired-end reads. We then determine a minimum path cover of DAG, where each path represents a haplotype. Therefore, the two nodes for which we computed DAG are divided as many times as paths were found. 

\begin{figure*}[t]
	\centering
	\includegraphics[width=0.7\textwidth]{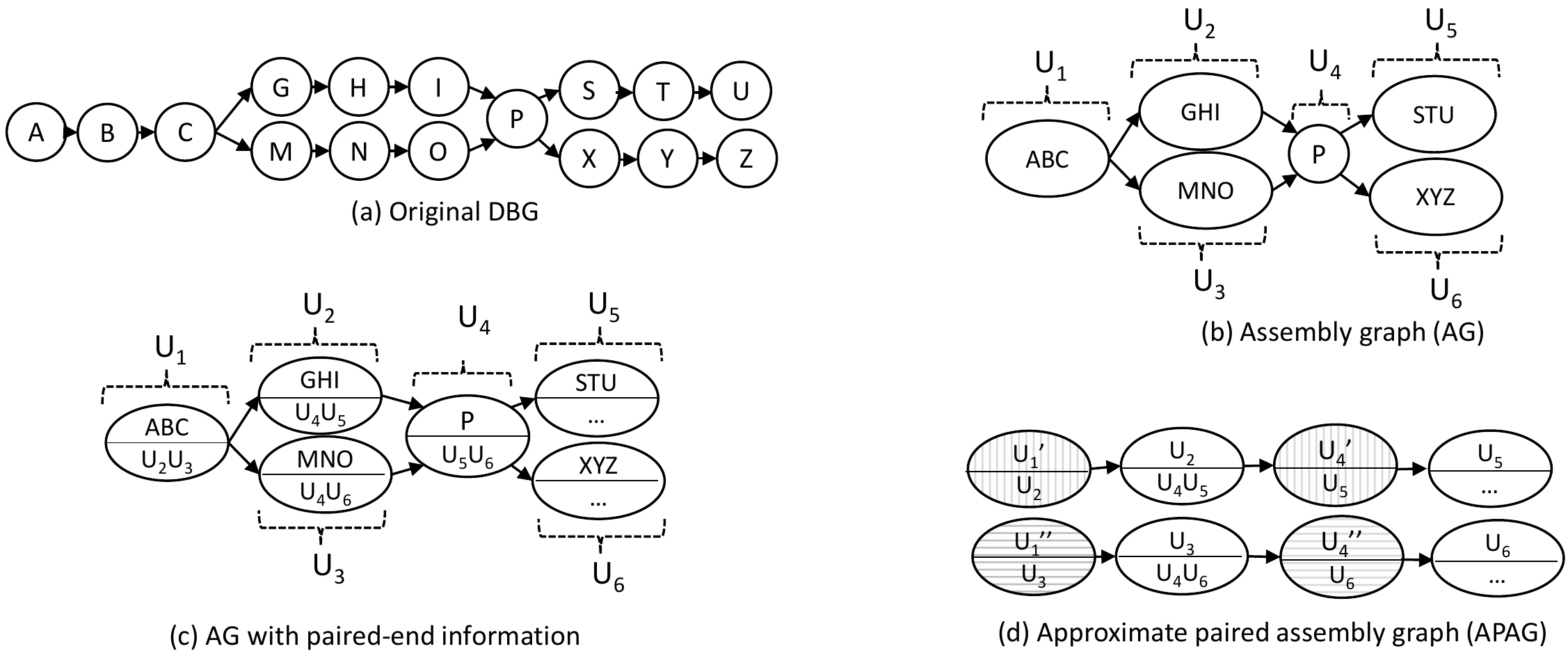}
	\caption{Example illustrating different steps of the proposed method from the original DBG to APAG. }\label{fig:assemblygraph}
\end{figure*}

For this to work properly, each DAG must be carefully processed to achieve a more reliable 
graph. This includes transforming each DAG into an offset flow network and solving a min-cost flow problem. With the detected flows, the DAG is corrected, yielding a more reliable graph. 
Finally,  the nodes of the AG are divided based on the path covers of its DAGs, and this results 
in a new AG, called approximate paired AG (APAG). Once this graph is polished, it is used to derive the contigs and their abundances. 
Figure \ref{fig:assemblygraph} shows a running example of this process, which will be used in this section to illustrate how to build the final APAG from the original DBG.

\subsection{Obtaining the AG}

First we will explain in detail how AG is produced. 

The nodes of AG are unitigs, i.e., compacted nonbranching paths of a DBG. Two unitigs are connected with an edge if the corresponding edge also exists in the DBG. 
AG is augmented with paired-end information by associating a set of paired unitigs $P(U)$ to each unitig $U$. The paired unitigs are inferred based on the paired-end reads. 

\subsubsection{Obtaining solid $k$-mers}

We denote a \kmer\ as \textit{genomic} if it appears in a haplotype of the sample and \textit{nongenomic} if it does not. Recall that,  in $D$, reads contain erroneous changes of base pairs, insertions, and deletions; thus,  there can be nongenomic \kmers. Therefore, we use the notion of {\em solid} \kmer, which denotes a \kmer\ that occurs at least $t$ times in the reads in $D$, where $t$ is an abundance threshold.  The selection of solid $k$-mers based on the  frequency of appearance \cite{euler} works well due to the high  coverage used in viral quasispecies assembly. Therefore, it is possible to establish a threshold to separate correct and erroneous information with high level of accuracy.
The process is simple; the entire read set is traversed, and those $k$-mers  whose frequency is above the threshold are classified as solid, whereas the rest are considered nongenomic.

There are several methods for selecting the threshold for solid \kmers. Chaisson and Pevzner \cite{chaisson2008short} approximated the number of reads covering a \kmer\ using a Poisson distribution. They selected a threshold such that few \kmers\ are expected to be covered by several reads below that threshold. 
Chaisson \textit{et al.} \cite{chaisson2009novo}  applied a Poisson and Gaussian mixture model to the $k$-mer frequency distribution, and the first local minimum is selected as the threshold.
Zhao \textit{et al.} \cite{zhao2010edar} used a clustering method based on the variable-bandwidth mean-shift algorithm.
In our recent study \cite{viaDBG}, we used a window over the \kmer\ frequency histogram to detect the threshold value where no frequency reduction is detected. 

\begin{figure}[t]
    \centering
 \subfigure[HCV-10]{
  \centering
  \includegraphics[width=.30\linewidth]{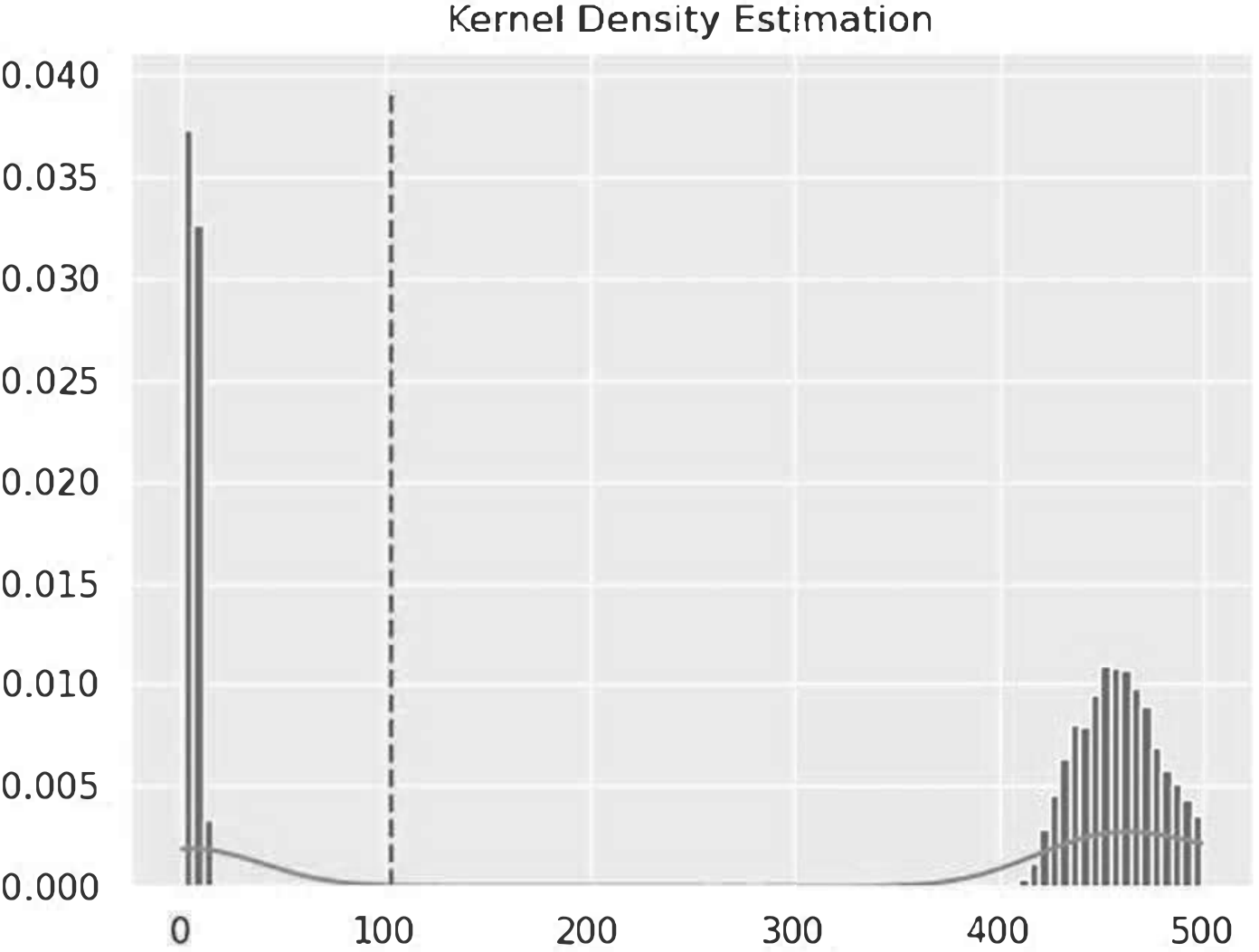}

  \label{fig:sub1}
}%
\subfigure[HIV-real]{
  \centering
  \includegraphics[width=.30\linewidth]{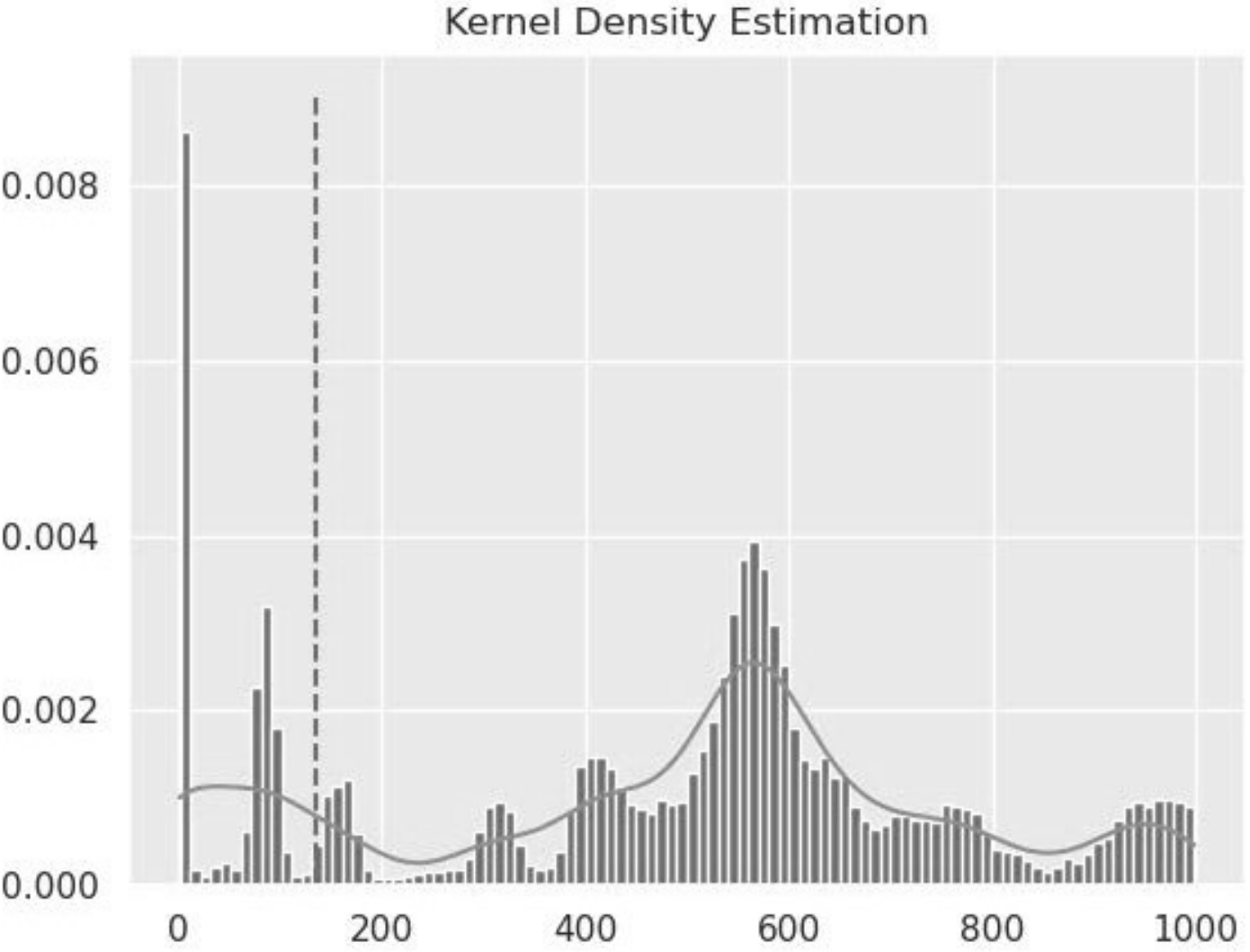}

  \label{fig:sub2}
}
\caption{Kernel density estimation of \kmer\ abundance for two datasets of the experimental evaluation (HCV-10 and HIV-real). We indicate with a dotted line the threshold selected for each of the datasets.}
\label{fig:test}
\end{figure}

We propose a new approach based on estimating the \kmer\ frequency density function through kernel density estimation using the Gaussian kernel. A brief introduction of kernel density estimation can be found in the supplementary material. 
The frequency is expected to decrease systematically, corresponding to nongenomic \kmers, up to a certain point where it stabilizes. Then, it subsequently begins to increase again, corresponding to genomic \kmers. Therefore, a good threshold would be the middle point between the minimum point and the previous maximum value of the function. 
The problem of possible local minima can be mitigated by increasing the suggested bandwidth; thus, producing an intended over-smoothing effect in the density function
, as shown in Figure \ref{fig:sub2}.

\subsubsection{Building and polishing the AG}

The solid \kmers\ obtained in the previous step are used to build a DBG. We  join all the nodes of the unitigs in just one node, as shown in Figure \ref{fig:assemblygraph}(b), to reduce the number of nodes of the graph. The resulting graph is called AG. We use BCALM2~\cite{bcalm} to construct AG.

First, the graph is polished by removing isolated unitigs and tips shorter than a given threshold. Then, we remove filigree edges, an approach previously employed in metagenomics \cite{metaspades}. Filigree edges are
the connections between unitigs that are not strongly supported. The idea is not to remove edges based on an overall threshold, which may remove edges corresponding to low-frequent haplotypes. Instead, edges in the graph are tagged as weak or strong based on the connection they support. For instance, the abundance of an edge $e$, $abu(e)$, denotes the number of reads that support the edge. Then, the abundance of a node $v$,  $abu(v)$, is defined as the maximum $abu(e)$ over all edges $e$ incident to $v$.
Then, every edge $e = (v_1,v_2)$ is tagged as weak  if $abu(e)*ratio < min(abu(v_1),abu(v_2))$ and strong otherwise. 
We set the $ratio$ to a default value of 5, lower than the typical value used in metagenomics, where abundances are lower. 

\subsubsection{Adding paired-end information to the unitigs}

Paired-end information is usually employed in a post-assembly step to join contigs into scaffolds such that the assembly can be extended or repetitive sections can be solved. However, this information has also been used in the previous stages of the assembly process. For instance, the  approximate paired DBG (APDB) \cite{pairedDBG} is  a modification of the DBG to use paired-end information directly in contig assembly. The same ideas were also employed for assembling  viral quasispecies \cite{viaDBG}. The aim is to use the paired-end information before building the contigs to untangle AG.
Thus, the paired-end information is included during the graph-building step. This addition requires time and increases space consumption. However, it does not represent a problem for viral datasets in practice.

More precisely, 
given a \kmer\  $k_d$, its paired \kmers\ are  $P(k_d)=\{k_g \mid k_g ~ \textrm{ is a solid $k$-mer and } \exists~ r_x \in D$  \textrm{ and a position $j$  such that } $L(r_x)[j\ldots j+k-1] =k_d \textrm { and } R(r_x)[j\ldots j+k-1] = k_g\}$.

\tolerance 10000 \pretolerance 10000
Let $K_y$ be a set of \kmers,  ${\mathcal U}(K_y)$ denotes the  unitigs where the \kmers\ in $K_y$  appear. Then, given a unitig $U_i$ formed by \kmers\ $k_1, k_2, \ldots, k_m$, its paired-end information is $P(U_i)=\{{\mathcal U}(P(k_1)), {\mathcal U}(P(k_2)), \ldots, {\mathcal U}(P(k_m))$\}. Therefore, to each vertex $U_i$ of the AG, we attach its $P(U_i)$. 

An example of paired unitigs is shown in the supplementary material. Figure \ref{fig:assemblygraph}(c) shows our AG with paired-end information. 

\subsection{Building the APAG}

Next, we will transform AG augmented with paired-end information to APAG. Similar to AG, the nodes of APAG are unitigs augmented with paired unitigs. However, in APAG, unitigs that occur in more than one haplotype will be replicated. Each instance of the unitig ideally corresponds to one haplotype and the paired unitigs in the same haplotype. The edges of APAG will connect the unitigs belonging to the same haplotype.

To construct APAG, we take each pair of adjacent nodes in AG and build a DAG, where its nodes are those two nodes and their paired unitigs. The DAG captures the connectivity between the unitigs in AG, by adding edges in DAG between the unitigs linked in AG through short paths. Therefore, each edge of DAG indicates that the connected nodes probably form part of the same haplotype. 
Then, we detect the haplotypes in that group of unitigs by finding paths in the DAG. We use that information to split the edge and the two adjacent nodes of AG according to the haplotypes found in DAG.
 
Therefore, DAGs are the core elements of the proposed method. This implies that they must be built with care to obtain reliable paths, including a correction based on a min-cost problem built on DAG.

\subsubsection{Processing adjacent nodes: building the DAGs}

Observe that 
if all occurrences of a unitig $U_i$ are due to a unique haplotype, then all paired unitigs of $U_i$ occur along a path in AG. Thus, they are all reachable from each other. 
 This occurs in our example of Figure \ref{fig:assemblygraph}(c) with, e.g., $U_2$, with paired unitigs $U_4$ and $U_5$. 
 We employ this information to detect haplotypes. For example, the pairing between $U_2$ and  $U_4$/$U_5$ highlights one of the two haplotypes in  Figure \ref{fig:assemblygraph}: $\{A,B,C,G,H,I,P,S,T,U\}=\{U_1,U_2,U_4,U_5\}$. The other haplotype is $\{A,B,C,M,N,O,P,X,Y,Z\}=\{U_1,U_3,U_4,U_6\}$.

However, if the unitig $U_j$ occurs in several haplotypes, the paired $k$-mers will span some site containing a mutation. In our example, $U_1$ appears in two haplotypes; thus, its paired information contains $U_2$ and $U_3$, which are not reachable from each other. However, the paired unitigs originating from the same haplotype are reachable from each other. We will use this reachability information to split the  nodes of AG into different nodes. Each node corresponds to a different haplotype; thus, producing the new APAG. The new APAG will be used to obtain the contigs.


Let $AG=(V_{ag}, E_{ag})$ be the AG. 
Our target is to divide the nodes in $V_{ag}$ into as many nodes as the number of haplotypes they belong to. Thus, we follow the next steps:

\begin{enumerate}
\item For each pair of adjacent nodes of $V_{ag}$, say $U_i$ and $U_j$,  a directed acyclic graph $\mathit{DAG}_{ij} = (V_{ij},E_{ij})$ is built, where
\begin{itemize}
    \item $V_{ij} = U_i \cup U_j \cup P(U_i) \cup P(U_j) $.
    \item There is an edge $e \in E_{ij}$ from $u \in V_{ij}$ to $v \in V_{ij}$, if and only if there is a path in  $AG$ from $u$ to $v$ that does not include any other node in $V_{ij}$ and has a length shorter than $2\Delta$, where $\Delta$ is the maximum error in the insert size. When searching for a path, we also limit the number of traversed branches by a threshold $T$ to ensure that the path search remains tractable. By default, we set $T$ to 10.
    
\end{itemize}

Observe that the  nodes of DAG are the two adjacent (in AG) unitigs $U_i$ and $U_j$, and their paired unitigs.   $U_i$ and $U_j$ overlap by $k-1$ bp\footnote{Recall that AG is a compaction of the DBG, so edges between unitigs are actually edges of the DBG, and thus the end of the source unitig overlaps $k-1$ bp with the beginning of the target unitig.}; thus, their paired unitigs (separated from $U_i$ and $U_j$, on average, by the insert size) should also be 
close by in $AG$ because their distance in $AG$ should not exceed $2\Delta$, twice the insert size error if they belong to the same haplotype.

\item Next, we will find the source to sink paths in DAG such that the paths correspond to the haplotypes. 
However, not all paths correspond to haplotypes. To identify the correct paths, we will use the coverage of the DAG nodes and edges along with the paired-end information of the reads. 

The coverage of the DAG edges is computed based on the coverage of the unitigs on the path in AG that the edge represents, where the coverage of a unitig is the average abundance of the $k$-mers of the unitig.
When computing the coverage of an edge $(u,v)$ of DAG, the process analyzes the paths in AG connecting them. It means that if there is only one path $p$, the  edges that reach nodes of  $p$ from  nodes that are not in $p$ and those coming from nodes in $p$ and reaching nodes that are not in $p$ correspond to haplotypes different from that containing $u$ and $v$. 
Thus, we analyze the paths connecting $U_i,U_j, u,$ and $v$. We traverse those paths. At each node, we add to a bag the coverages of edges coming from nodes that do not belong to $p$. Then, we distribute those \textit{excess} coverages in the bag among the nodes forming a fork when we find one. Therefore, we combine the coverage and paired-end information of the reads.

 The final coverage is that of the last edge of $p$ minus the incoming coverages that were not assigned to  nodes leaving $p$. The exact method for computing the coverages is described in more detail in the supplementary material.

 Once the edge and node coverages are computed, we want to find a set of paths in DAG and associated abundances to explain the coverage of edges and nodes.

Figure \ref{fig:dag} shows the process of building a reliable DAG for detecting the haplotypes that pass through the adjacent nodes $U_2$ and $U_4$. Figure \ref{fig:dag}(a) shows AG of our running example with the coverages of nodes and edges, i.e., the number of reads supporting those nodes and edges. 
Figure \ref{fig:dag}(b) shows the first version of DAG $\mathit{DAG}_{24}$. Observe that
there could be two paths $U_2$, $U_4$ and $U_5$, and $U_2$, $U_4$, and $U_6$; thus, we must check if they are correct using the coverages of edges in AG and the paired information.
Figure \ref{fig:dag}(b) shows that, even after applying the previous method for computing the coverage of the edges of DAG, the coverages may not be consistent.  
Observe that  $U_4$ has two exits, adding a total output coverage of 12. This is not possible since $U_4$ only receives an input coverage of 10.

\begin{figure*}[t]
	\centering
	\includegraphics[width=0.7\textwidth]{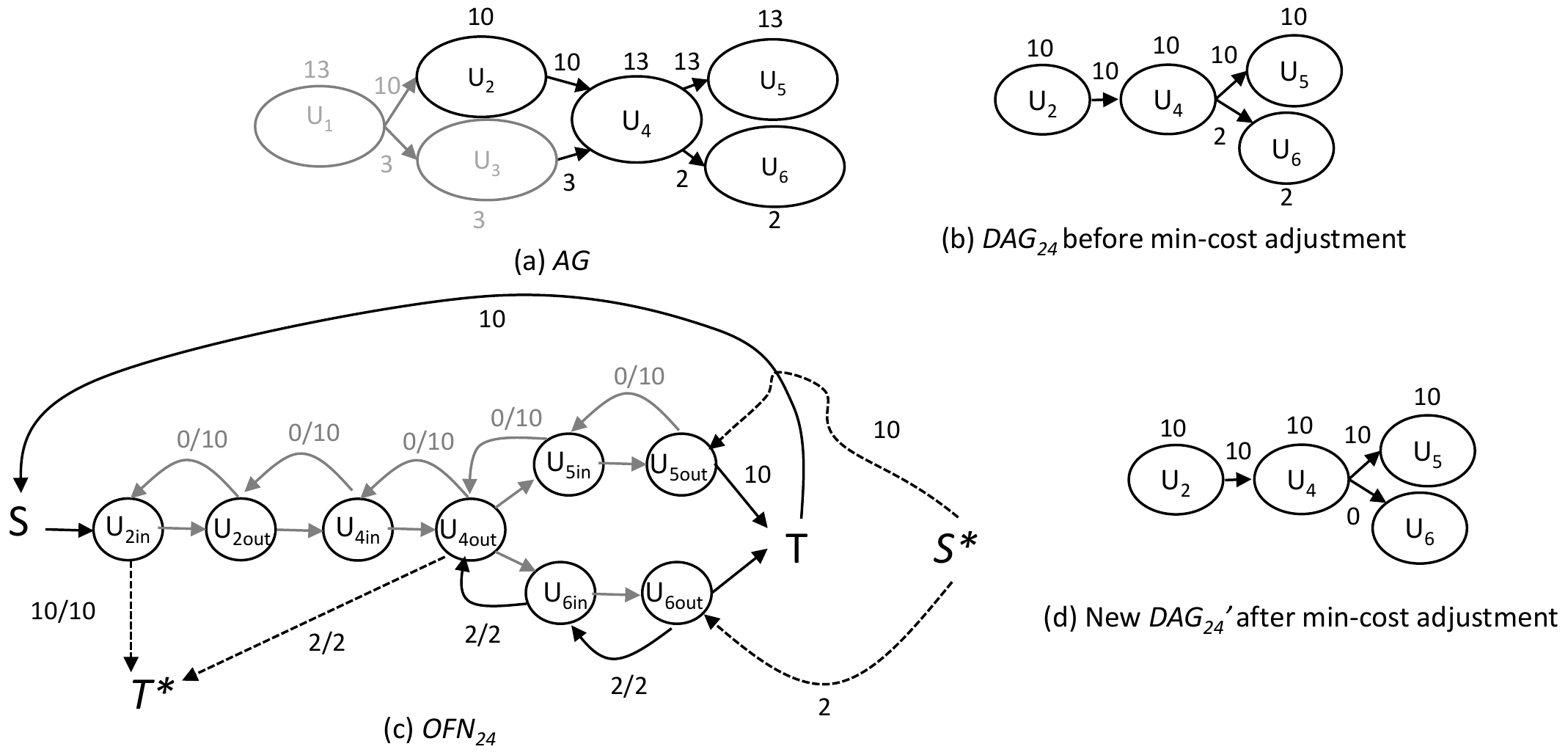}
	\caption{Example of the construction and coverage adjustment of DAG corresponding to the adjacent nodes $U_2$ and $U_4$. We show the original AG, $\mathit{DAG}_{24}$, $OFN_{24}$, and $\mathit{DAG}_{24}'$ 
	}\label{fig:dag}
\end{figure*}

\end{enumerate}

Therefore, the coverages of the graph are inconsistent. Moreover,  there can be false edges due to false paths in $AG$. Their origin is due to the shared $k$-mers between different but similar haplotypes.
Additionally, there can be several real edges and thus paths, but the observed coverages may not be consistent with each other.

To solve these problems, we employ the coverage of the unitigs (nodes of DAG), i.e., the number of reads supporting them and their connections.

We need to assign a flow to the graph to satisfy the \textit{conservation} property of the flow.

It means that the amount of flow reaching each node must be the same as those leaving the node.  A simple heuristic, such as taking first the heaviest paths, will work accurately with a graph in that form.
However, conservation is not enough; the flow assigned to each node/edge must be as close as possible to the observed coverage.

Our example, $\mathit{DAG}_{24}$, is simple, but in a very complex graph, a decision on the flow distribution must be taken at each fork; however, the flow that reaches the fork might depend on several previous decisions. Therefore, the optimal solution is very complex. We will use a min-cost flow that will solve the two problems using a cost function, i.e., it will modify DAG such that i) satisfies the conservation law, ii) the new flow assigned to the edges  is  close  to the observed coverage.

To assign a flow minimizing the difference from the observed coverage of the unitigs in DAG, the min-cost flow uses a cost function that increases when the difference grows. Therefore, we obtain our solution by minimizing that cost. The main advantage of this method is that it decides at all forks using a  polynomial-time algorithm, minimizing the difference between the assigned flow and observed coverage \textit{globally}. It means that it does not compute the best cost in the fork but computes the cost as the sum of costs at all edges of the graph to decide the flow in each edge of a fork. 

With the corrected graph, the next step is to decompose the flow into paths in the graph in the most parsimonious way. Finally, the corrected flow and its decomposition into paths will give us the relative abundance of the haplotypes. 

Consider the following: 
let  $cov(v)$ be the coverage of the nodes $v \in V_{ij}$,  $cov(u,v)$ be  the coverage of the edges $(u,v) \in E_{ij}$, and $c(y)$ a cost function that can be applied  to  nodes $v\in V_{ij}$  or  edges  $(u,v) \in E_{ij}$. Following the approach proposed by \cite{traph}, we use, for our problem,
the same cost function for nodes and edges $c(x) = x^2$. Here, $x$ is the error in the frequency estimation for nodes and edges.

The objective is to find the set of paths ${\mathcal P}_{ij}$ from the sources of $\mathit{DAG}_{ij}$ to the sinks of $\mathit{DAG}_{ij}$ that minimizes:

\medskip
\(\begin{array}{ll}
    err =& \sum_{u \in V_{ij}}c_{u}\big(|cov(u) - \sum_{p\in {\mathcal P}_{ij}|u \in p} a(p)|\big)+\\&\sum_{(u,v) \in E_{ij}}c_{uv}\big(|cov(u,v) - \sum_{p\in {\mathcal P}_{ij}|(u,v) \in p} a(p)|\big),
\end{array}\)
\medskip

\noindent
where  $a(p)$, for each path $p \in {\mathcal P}_{ij}$, is the  estimate abundance level of the path obtained by the solution of the min-cost flow problem and the decomposition of the flow into paths. 



\subsubsection{Translating DAGs into offset flow networks}

In this section, we explain how we use the min-cost network flows to first determine an optimal flow in DAG and then split the flow into paths. A brief introduction to network flows can be found in the supplementary material.

To determine such an optimal flow, we
translate DAGs into offset flow networks and solve the min-cost problem in those offset flow networks. The min-cost flow in the offset flow network models how the coverages in DAG need to be modified to create an optimal feasible flow.

For each $\mathit{DAG}_{ij}=(V_{ij},E_{ij})$, we build the offset flow network $\mathit{OFN}_{ij}$ as follows:
\begin{itemize}
    \item Replace every node $v \in V_{ij}$ by two nodes $v_{in}$ and $v_{out}$ and two arcs: $(v_{out},v_{in})$, with capacity $b(v_{out},v_{in})= cov(v)$ and cost function $c_{v_{out},v_{in}}(x)=c_{v}(x)$, and $(v_{in},v_{out})$ with infinite capacity and the same cost function. The flow on the arc $(v_{out},v_{in})$ will be nonzero if the coverage of $v$ in DAG needs to be decreased in the optimal solution. However, the flow on the arc $(v_{in},v_{out})$ will be nonzero if the coverage of $v$ needs to be increased.

\item Replace the arc between each pair of original nodes  $(u,v)\in E_{i,j}$  by two edges: $(v_{in},u_{out})$ with capacity $b(u,v)=cov(u,v)$, and cost function $c_{v_{in},u_{out}}(x)=c_{u,v}(x)$, and $(u_{out},v_{in})$ with infinite capacity, and the same cost function.  Similarly, a nonzero flow on the forward arcs indicates that the coverage of the arc $(u,v)$ should be increased. In contrast, a nonzero flow on the backward arcs indicates that the coverage should be decreased.

\item Add start ($S$) and target ($T$) nodes. For each source node $s$ in $\mathit{DAG}_{ij}$, add an arc from $S$ to $s$ with  cost $c_{S,s}(x) = 0$ and infinite capacity. Similarly, for each target node $t$ in $\mathit{DAG}_{ij}$, add an arc from $t$ to $T$ with $c_{t,T}(x) = 0$ and infinite capacity. Source nodes are the nodes with no incoming arcs, and target nodes are the nodes with no outgoing arcs.

\item  Add an artificial source ($S^*$) and sink ($T^*$) to manage the exogenous flow and initialize the exogenous flow of each node of $OFN_{ij}$ to $q_v=0$. These will be used to balance the flow on nodes where the total coverage of the incoming arcs and the total coverage of the outgoing arcs do not match. 

\item  For any node $v$ with $$q_v=\sum_{u \in V} cov(v,u)-\sum_{u \in V} cov(u,v) < 0$$  add an arc from $S^*$ to the node $v$ with capacity set to $q_v$ and cost $c_{S^*,v}(x) = 0$. Update the exogenous flow of $S^*$: $q_{S^*} = q_{S^*} + q_v$. 
\item  For any node $v$ with $$q_v=\sum_{u \in V} cov(v,u)-\sum_{u \in V} cov(u,v) > 0$$  add an arc from $v$ to $T^*$ with capacity set to $q_v$ and cost $c_{v,T^*}(x) = 0$. Update the exogenous flow of $T^*$: $q_{T^*} = q_{T^*} + q_v$. 
\item  Add an edge from $T$ to $S$ to close the offset network with cost $f(T,S)=0$ and infinite capacity. 
\end{itemize}

Figure \ref{fig:dag}(c)  shows the resulting network flow for the offset network of  $\mathit{DAG}_{24}$ of Figure \ref{fig:dag}(b). On the edges, the label $x/b$  means that $b$ is the  node's capacity, and $x$ is the flow that passes through it when solving the min-cost problem. Edges with only a value $x$ means that the edge has $\infty$ capacity, and the flow passing through it is $x$.

\subsubsection{Adjusting DAGs according to the flow in the offset flow network}

Once the min-cost problem  in the offset flow network is solved, the original DAG is modified as follows.

Let us consider $\mathit{DAG}_{ij}$ and its $OFN_{ij}$.
Let $cov(u,v)$ be the coverage of the edge $(u,v) \in \mathit{DAG}_{ij}$, $x_{u_{out} v_{in}}$ be the flow  through the edge $(u_{out},v_{in})$ in $OFN_{ij}$, and $x_{u_{in} v_{out}}$ be the flow  through the edge $(u_{in},v_{out})$ in $OFN_{ij}$. Then, the new coverage  $cov(u,v)'$ of the edge $(u,v)$ in the modified $\mathit{DAG}_{ij}'$ is: 
$cov(u,v)'=cov(u,v)-x_{u_{in}v_{out}}+x_{u_{out} v_{in}}$.

Observe in Figure \ref{fig:dag}(d) the new $\mathit{DAG}_{24}'$, the output flows of $U_4$ match the input flows.

\subsubsection{Obtaining paths in the new DAGs}

For each  adjusted $\mathit{DAG}_{ij}'$,  we build a set of paths ${\mathcal P}_{ij}=p_1,p_2,$ $p_3,\dots,p_p$ from the sources to the sinks of $\mathit{DAG}_{ij}'$ with weights $w_1,w_2,w_3,\dots,w_p$ satisfying:
\begin{equation*}
    cov(u,v)' = \sum_{\forall p_t \in \mathcal P_{ij}|u \in p_t~and~v \in p_t}w_t
\end{equation*}
\noindent
for all edges $(u,v)$ of $\mathit{DAG}_{ij}'$.

Analogously to prior research to solve RNA-seq problems \cite{traph}, we are interested in {\em parsimoniously} explaining the abundance distribution over the sample. We are interested in finding the set of paths ${\mathcal P}_{ij}$  with the lowest number of paths. 
A well-known result is that decomposing a flow into a minimum number of paths is an NP-hard problem in a strong sense. Therefore, only heuristic methods can be run in a reasonable time. In this work,  we use a mixture of two  heuristics to build the final set of paths.

We iteratively find the source to sink paths in the graph until the paths are a decomposition of the flow. We resolve branches primarily using paired-end information and secondarily choosing the heaviest path when the paired-end information is unavailable or does not point to a unique solution. To achieve this, we proceed in two steps as follows:

    \begin{itemize}
        \item For each node $u \in V_{ij}$, we store the list of nodes in $V$ such that $u$ is part of their paired-end information, i.e., $paired\_with[u] = \{v_1,\dots,v_n\}$ iff $ u\in P(v_i), i = 1,\dots,n $.
        \item When traversing the graph to find a path, we remember the parent of the last node on the path with an indegree greater than one. We call this saved node haplotype reference node. When we find a node with an outdegree greater than one, we must decide which branch to follow. Then, we take the haplotype reference node and employ the map built in the previous step.
              There are three possibilities: none of the branches of the fork is pointed by any node in the paired-end information of the haplotype reference node, then we use the heaviest path heuristic; only one branch of the fork is pointed, then we follow that path; more than one branch is pointed, then we follow the heaviest path of the pointed branches.
    \end{itemize}

In the example of Figure \ref{fig:dag}(d), the only resulting path is $p_1= U_2, U_4, U_5$, with weight 10. We have discarded the path $U_2, U_4, U_6$, with weight 0, which, if not removed, would be responsible for mixing haplotypes.  

Additionally, in DAGs, we have information about the relative abundance of the haplotypes.

\subsubsection{Polishing the paths} 
It is necessary to solve one min-cost flow problem for each pair of adjacent nodes in AG. Ideally, the number of paths reported for each pair coincides with the number of haplotypes to which the pair belongs. However, this does not happen in practice due to, for instance, shared regions or coverage inconsistencies; thus, a wrong number of paths might be reported. Therefore, a soft polishing is performed to overcome this problem. This polishing involves tasks, such as removing paths that do not contain both nodes under study, paths with suggested flow below the $k$-mer solid threshold, and paths with low simultaneously support from the given nodes.

The expected output once the polishing has been completed is to have for each pair of nodes both the haplotypes it belongs to and the estimation of the abundance of each haplotype given by the suggested flow.

\subsubsection{Building the APAG}

For each pair of adjacent nodes  $U_i$ and $U_j$ in $AG$, we take the set of paths $\mathcal{P}_{ij}$, and for each path $p \in \mathcal{P}_{ij}$, unless they are already in APAG, we create the nodes:

\begin{itemize}
\item ${U_i}_{P(U_i) \cap p}$, which is the node with label $U_i$ with paired information $P(U_i) \cap p$.

\item ${U_j}_{P(U_j) \cap p}$, which is the node with label $U_j$ with paired information $P(U_j) \cap p$.

\end{itemize}

In our example, $\mathit{DAG}_{24}$ produced only one path ${\mathcal P}_{24}=\{p_1\}$, and $p_1=\{U_2, U_4, U_5\}$. Then, we create two nodes:

\begin{itemize}
    \item ${U_2}_{P(U_2) \cap p_1}$. Since $P_{U_2}=U_4, U_5$, we create the node labeled $U_2$ in Figure \ref{fig:assemblygraph}(d) with paired information $U_4, U_5 $. 
    
    \item ${U_4}_{P(U_4) \cap p_1}$. Since $P_{U_4}=U_5, U_6$, we create the node labeled $U_4'$ in Figure \ref{fig:assemblygraph}(d), with paired information $U_5$. 
    
\end{itemize}

As shown in Figure \ref{fig:assemblygraph}(d),   \textit{APAG} of our running example and our nodes $U_2$ and $U_4'$ are in the upper haplotype; the remaining nodes of the haplotype are computed from DAGs $\mathit{DAG}_{12}$ and $\mathit{DAG}_{45}$. 

The nodes $U_4'$ and $U_4''$ correspond to the same unitig, but those nodes have different paired information, meaning that they correspond to  different haplotypes.

\subsection{Extracting the haplotypes and their abundances}

Finally, we polish APAG by removing isolated nodes, which can be wrongly built when splitting one node more than necessary. We also remove  short tips and paths with lengths below 500 bp, which are not expected to be correct contigs. After polishing APAG, the only remaining step is to traverse it and report the correct paths. 

For this, we solve  a min-cost flow problem on APAG. APAG is transformed into an offset network using the same method used previously for DAGs. Once the offset network is built, a min-cost flow is solved on it using cost function $c_{uv}(x) = (x - cov(u,v)')^2$. Therefore, the first paths to be flooded will be the ones with more flow capacity. Then, we obtain the haplotypes and their abundances by decomposing the flow into weighted paths using a greedy algorithm to extract the heaviest path.

A similar approach has been previously introduced by VG-Flow, which solves a min-cost flow problem over a variation graph, using the number of paired-end reads mapping a particular edge as edge labels. Then, it performs a greedy path extraction to obtain the set of paths $P$ and solves a linear programming problem to polish the results. In contrast, ViQUF skips the variation graph using APAG. It also avoids mapping the reads  using the suggested flows as labels for the edges. 
Although \ours\ could obtain the haplotype abundance estimation based on the estimated flows in APAG, the estimation for some haplotypes may not be accurate due to the lack of precision on the abundances for heads/tails plus some wrong estimations that can appear on the split computation. Therefore, like VG-Flow, \ours\ performs a flow polishing based on a simple linear programming problem. Given $APAG = G = (V,E)$ a graph, $P = \{p_1, p_2, \dots,p_n\}$ a set of paths on $G$, and $f(p_i)$ a flow over the paths, we define
\begin{equation}
\begin{aligned}
\min_{x} \quad & \sum_{u \in V} \big| ab_u - \sum_{p_i, u \in p_i} f(p_i)x_{p_i}\big|\big(\frac{L_u}{k}- 1\big)\\
\textrm{s.t.} \quad & x=[x_{p_1},\ldots ,x_{p_n}] \in R^n_{\geq 0}\\
\end{aligned}
\label{equ:min}
\end{equation}
where $L_u$ and $ab_u$ are the length and abundance of the unitig, respectively; $k$ is the $k$-mer size, and $f(p_i)x_{p_i}$ is the flow of the path $p_i$ after polishing. Equation \ref{equ:min} minimizes the accumulated difference for each vertex and the assigned flow for each path it belongs to. The second term of the expression is a correction factor that gives a higher weight to the longest; thus, the most reliable, unitigs. The aim is to obtain, for each path, a weight $x_{p_i}$ that minimizes the difference between the abundance of each unitig and the given flow. Although simple, it is quite effective in practice and, even for complex cases, the frequency estimation improves significantly compared to the previous estimation.

\subsection{Theoretical complexity analysis}


 To build the initial AG, we must build and traverse a DBG from the reads. We use GATB library \cite{gatb} for this step, which takes $\mathcal{O}(N)$ time, where $N$ denotes the total length of all reads.


 To build APAG, we need to associate each unitig with the unitigs through the paired-end information. To do so, we traverse the read set and, for each pair of $k$-mers in the left and right reads, we locate the corresponding unitigs and add the right unitig to the set of paired unitigs of the left unitig. Assuming constant time access to the unitigs using a hash function, this takes $O(N)$ time. 

To build $\mathit{DAG}_{ij}$ for two nodes $U_i$ and $U_j$ of $AG$, we traverse $AG$ from the first node $U_i$ until we reach a specific distance  (we use $2\Delta$ by default)  without finding any node from $pe = P(U_i) \cup P(U_j)$. In the worst case, the number of iterations of this step is $n_h(k+2\Delta)$, where $n_h$ is the number of haplotypes. This leaves us a complexity of $\mathcal{O}(n_h(k+2\Delta))$.

After building DAG, it is polished. For this, DAG is traversed. This costs  $\mathcal{O}(|V_{ij}|+|E_{ij}|)$, where $|E_{ij}|$ and $|V_{ij}|$ are the number of edges and vertices of $\mathit{DAG}_{ij}$, respectively. Furthermore, in the worst case $|V_{ij}| = n_h(k+2\Delta)$, indicating that the insert size is maximum and every \kmer\ within the insert size belongs to $pe$. In DAGs, the $|E_{ij}|$ is bounded to $|V_{ij}| \choose 2$. However, even in the worst case scenario, this is impossible, and the upper limit is $4|V_{ij}| = 4n_h(k+2\Delta)$, meaning that each node is at most connected to four other nodes. Again, the complexity of this step is also $\mathcal{O}(n_h(k+2\Delta))$.

 The third and hardest step is solving the min-cost flow problem. There are two possibilities: if the cost function is concave or convex. If it is concave, the problem is NP-hard; however, the problem can be solved in polynomial time if the cost function is convex \cite{genome_scale}. Given a circulation problem on a network $N = (G,l,u,c)$, where $G$ is a directed graph, $l$ and $u$ are non-negative functions for demand and capacity for every arc, respectively, and $c$ is a convex cost function, the minimum-cost circulation, flow, problem can be solved through cycle-canceling algorithm in $\mathcal{O}(nm^2CU)$, where $n = |V(G)|$, $m = |E(G)|$, $C$ is the maximum value of the costs, and $U$ is the maximum capacity in an edge. As described earlier,  $n \leq n_h(k+2\Delta)$ and $m \leq 4n$; thus, the algorithm runs in $\mathcal{O}(n^3CU) = \mathcal{O}((n_h(k+2\Delta))^3CU)$. 
 
 Finally, building and polishing DAG and solving the min-cost flow problem is repeated for each edge or pair of adjacent nodes in AG, which is four times the number of $k$-mers in the sample 
 in the worst case scenario. Thus, the theoretical complexity of these steps is $\mathcal{O}((N)(n_h(k+2\Delta))^3CU)$.

The final step is to solve a minimum-cost flow over APAG, which in the worst case is the same as AG, which in the same scenario is DBG. Therefore, we can use the same previous complexity $\mathcal{O}(nm^2CU)$, but in this case $n \leq N$ and $m \leq 4n = 4(N)$, resulting in a complexity of $\mathcal{O}(N^3CU)$.

As summary, the entire algorithm takes $O((N) (n_h(k+2\Delta))^3CU + (N^3CU))$. 
If we assume $k$ and $\Delta$ are constants, we obtain $\mathcal{O}(N^3CU)$. According to the theoretical analysis, building APAG and solving the min-cost flow in APAG are the most time-consuming steps. However, building APAG takes the longest in practice.


\section{Experimental evaluation}\label{sec:experiments}

We compare ViQUF with the most recent solutions for \textit{de novo}  haplotype-aware full-length viral quasispecies assembly: Virus-VG \cite{virusvg} and VG-Flow \cite{vgflow}. We also include in this comparison viaDBG \cite{viaDBG}, a related solution for {\em de novo} assembly of viral quasispecies, and PEHaplo \cite{pehaplo} (which do not provide haplotype abundance estimations). Additionally, we compare ViQUF with two reference-based viral quasispecies construction methods: CliqueSNV~\cite{cliquesnv} and PredictHaplo \cite{predicthaplo}. About parameters selections, we followed the recommendations given for all the authors in their papers. For ViQUF, we set the $k$-mer size to 121.

\subsection{Benchmarking data}

In our experimental evaluation, we used simulated and real MiSeq sequencing data. We employed the methodology and datasets used by related work \cite{savage,virusvg,vgflow}, which are described as follows.

\textit{Real data with ground truth.}
We used a gold standard benchmark for viral assembly \cite{Giallonardo14}. The reads were produced from five HIV haplotypes using Illumina MiSeq (2$\times$250 bp with an error of about 0.3\%) with 20000x coverage. Since the five haplotypes in the sample are known, it is possible to validate the achieved results. Table \ref{table1} presents the main characteristics of this dataset (HIV-real).

\textit{Synthetic benchmarks.}
Four simulated datasets were used, consisting of 2$\times$250 bp Illumina reads from different viruses: Human immunodeficiency virus (HIV), Poliovirus (POLIO), Hepatitis C virus (HCV), and Zika virus (ZIKV). These datasets correspond to those employed by \cite{vgflow} in their experiments. We denote these datasets by HIV-5,  POLIO-6, HCV-10, and ZIKV-15, respectively.
Table \ref{table1} also shows the main characteristics of these datasets.
\begin{table}
	\caption{Main characteristics for the datasets used in the experiments.}\label{table1}
	\scriptsize
	\setlength{\tabcolsep}{2pt}
	\begin{center}
		\begin{tabular}{|l|l|c|c|c|r|r|}
			\hline
			& Virus  & Genome  & Average  & \# Haplo- & Abun- & Diver- \\
			&  Type &  Length (bp) &  Coverage &  types & dance & gence \\\hline
			HIV-real & HIV-1 & 9487--9719 & 20000$\times$ & 5 & 10\%--30\% & 1\%--6\% \\ \hline
			HIV-5 & HIV-1 & 9487--9719 & 20000$\times$ & 5 & 5\%--28\% & 1\%--6\% \\ \hline
			ZIKV-15 & ZIKV & 10251--10269 & 20000$\times$ & 15 & 1\%--13\% & 1\%--12\% \\ \hline
			HCV-10 & HCV-1a & 9273--9311 & 20000$\times$ & 10 & 5\%--19\% & 6\%--9\% \\ \hline
			POLIO-6 & Poliovirus  & 7428--7460 & 20000$\times$ & 6 & 1.6\%--51\% & 1.2\%--7\% \\ \hline
		\end{tabular}
	\end{center}
\end{table}

\subsection{Evaluation metrics}

For all datasets, there are ground truth sequences. Thus, we evaluate the assemblies by comparing the obtained contigs against these. We used MetaQUAST \cite{metaquast} for this evaluation. MetaQUAST gives multiple alignment statistics and an overall assembly evaluation. We used MetaQUAST with the option ``--unique-mapping'' meaning that each contig can map to only one haplotype. For each dataset, we report genome fraction retrieved, N50, and errors (misassemblies and percentage of mismatches/indels/Ns). The genome fraction is defined as the percentage of the target haplotypes contained in the set of the obtained contigs (all contigs with lengths larger than 500 bp). We used N50 to measure the fragmentation level of the assembly. The N50 is defined as the length of the shortest contigs in the assembly of at least the length of half of the total assembly.
We show misassemblies to measure the behavior on repetitive sections, and error rate, which is calculated as the sum of mismatch rate, indel rate, and N-rate, measures the correctness of the assembly, showing if a set of contigs is correct or just a random graph extension. The supplementary material includes the results for the measures of precision and recall, following the extension proposed by \cite{PMID:32151775} for {\em de novo} methods.

\subsection{Performance analysis}

\begin{table*}[]
\caption{Results for the four viral quasispecies assembly tools. We show percentage genome (the fraction of all haplotypes retrieved by each method), N50 (the length of the shortest contig needed to be included to cover at least half of the total assembly), the number of misassemblies, the error rate of the assembly (the sum of mismatch rate, indel rate, and N-rate), elapsed time, peak memory usage, mean estimated error (MEE) of haplotype frequencies, and the standard quasideviation of the estimated error of haplotype frequencies ($\hat{S}_{EE}$). 
For Virus-VG and VG-Flow, we show the elapsed time and memory usage separated into the time for contig assembly using SAVAGE (first value) and for the full haplotype reconstruction (second value). Similarly, for the reference-based methods, PredictHaplo and CliqueSNV, we show the elapsed time and memory usage to align the reads to the reference (first value) and run the method (second value) separately. }\label{table3}
\centering
\footnotesize
\begin{tabular}{l|l|c|r|c|c|r|r|c|c|}
\cline{2-10}
&& \multicolumn{1}{c|}{\multirow{2}{*}{\% Genome}} & \multicolumn{1}{c|}{\multirow{2}{*}{N50}} & \multicolumn{1}{c|}{misass-} & \multicolumn{1}{c|}{\% error} &\multicolumn{1}{c|}{elap time} & \multicolumn{1}{c|}{memory} & \multicolumn{1}{c|}{MEE} & \multirow{2}{*}{$\hat{S}_{EE}$}\\

dataset& method  & \multicolumn{1}{c|}{}  & \multicolumn{1}{c|}{}& emblies & rate & \multicolumn{1}{c|}{(min)} & \multicolumn{1}{c|}{(GB)} &  \multicolumn{1}{c|}{(\%)} & \multicolumn{1}{c|}{}\\\hline

\multicolumn{1}{|l|}{\multirow{7}{*}{HCV-10}} & Virus-VG & 99.30\% & 9231   & 0 & 0.002 & 913.48 + 1009.08 & 26.13 | 8.35 & 0.05 & 0.04 \\ \cline{2-10}   
\multicolumn{1}{|l|}{} & VG-Flow & 99.79\% & 9293   & 0& 0.001&  913.48 + 559.56 & 26.13 | 8.29 & 0.05 & 0.04 \\ \cline{2-10} 
\multicolumn{1}{|l|}{} & viaDBG & 97.72\%  & 8934   & 0   & 0.005&  69.10& 2.81 & - & - \\ \cline{2-10} 
\multicolumn{1}{|l|}{} & PEHaplo & 94.78\% & 8661  & 0 & 0.013 & 68.45 & 8.94 & - & -  \\ \cline{2-10}
\multicolumn{1}{|l|}{} & PredictHaplo & 89.79\% & 9273 & 0 & 0.044 & 4.11+175.73 & 0.08 | 1.14 & 6.74 & 6.44  \\\cline{2-10}
\multicolumn{1}{|l|}{} & CliqueSNV & 9.97\% & 9273 & 0 & 2.100 & 4.11+3494.09 & 0.08 | 17.24 & 17.6 & 25.10  \\\cline{2-10} 
\multicolumn{1}{|l|}{} & \ours\   & 97.37\% & 8911  & 0& 0.008   & 3.51  & 1.09 & 0.15 & 0.13  \\ \hline\hline

\multicolumn{1}{|l|}{\multirow{7}{*}{HIV-5}}& Virus-VG & 96.85\% & 9632 & 2 & 0.332 &  1619.34 + 312.68 & 26.83 | 0.64 & 6.12 & 5.57  \\ \cline{2-10}
\multicolumn{1}{|l|}{} & VG-Flow & 96.87\% & 9625 & 2 & 0.331 & 1619.34 + 312.20 & 26.83 | 0.65& 5.25 & 4.79\\ \cline{2-10} 
\multicolumn{1}{|l|}{} & viaDBG  & 97.50\%& 8046& 2& 0.151 &  62.34  & 2.89 & - & -  \\ \cline{2-10} 
\multicolumn{1}{|l|}{} & PEHaplo & 78.59\% & 9328  & 2 & 0.690   & 73.33 & 4.84 & - & -  \\ \cline{2-10}
\multicolumn{1}{|l|}{} & PredictHaplo & 99.90\% & 9663 & 0 & 0.591 & 4.00+120.13 & 0.09 | 1.05 & 6.26 & 5.59  \\\cline{2-10}
\multicolumn{1}{|l|}{} & CliqueSNV & 99.86\% & 9649 & 0 & 1.152 & 4.00+93.67 & 0.09 | 8.51 & 7.02 & 4.06  \\ \cline{2-10}
\multicolumn{1}{|l|}{} & \ours\  & 99.71\% & 9237 & 2& 0.321  & 3.26 & 1.07 & 3.09 & 1.88  \\\hline\hline

\multicolumn{1}{|l|}{\multirow{7}{*}{POLIO-6}}& Virus-VG & 89.96\% & 7436 & 0 & 0.141 & 3455 + 201.23 & 17.30 | 0.73 & 1.56 & 1.18 \\ \cline{2-10}
\multicolumn{1}{|l|}{} & VG-Flow & 99.49\% & 7388 & 2 & 0.137 & 3455 + 532.33 & 17.30 | 0.30 & 2.18 & 2.48\\ \cline{2-10} 
\multicolumn{1}{|l|}{} & viaDBG & 73.81\%& 1760 & 0 &  0.018  & 49.21  & 2.52& - & -   \\ \cline{2-10} 

\multicolumn{1}{|l|}{} & PEHaplo & 98.15\% & 7428 & 0 & 0.125 & 107.96 & 3.63 & - & -  \\\cline{2-10}
\multicolumn{1}{|l|}{} & PredictHaplo & 49.81\% & 7428 & 0 & 0.646 & 3.80 + 82.35 & 0.10 | 0.92 & 11.08 & 11.65  \\\cline{2-10}
\multicolumn{1}{|l|}{} & CliqueSNV & 83.07\% & 7428  & 0 & 1.844 & 3.8 + 27.95 & 0.10 | 8.45 & 4.74 & 5.27  \\\cline{2-10}
\multicolumn{1}{|l|}{} & \ours\   & 97.40\% & 7428 & 0 & 0.247 & 2.61 & 1.06 & 3.05 & 1.79 \\ \hline\hline

\multicolumn{1}{|l|}{\multirow{7}{*}{ZIKV-15}}  & Virus-VG & 99.56\% & 10212   & 0 & 0.077 & 706 + 407.51  & 13.45 | 1.37 & 0.94 & 0.70 \\ \cline{2-10}
\multicolumn{1}{|l|}{} & VG-Flow & 83.05\% & 10210 & 0 & 0.144 & 706 + 406.22 & 13.45 | 0.62& 2.00 & 1.85\\ \cline{2-10} 
\multicolumn{1}{|l|}{} &  viaDBG  & 89.85\% & 1398 & 0& 0.110 & 65.48 & 3.25 & - & -  \\ \cline{2-10} 

\multicolumn{1}{|l|}{} & PEHaplo & 98.32\% & 10247 & 0 & 2.05 & 321.53 &  8.80 & - & -  \\ \cline{2-10}
\multicolumn{1}{|l|}{} & PredictHaplo & 46.65\% & 10251 & 0 & 0.133  & 4.06+149.68 & 0.08 | 1.11 & 6.33 & 7.68 \\\cline{2-10}
\multicolumn{1}{|l|}{} & CliqueSNV & 66.66\% & 10251 & 0 & 0.036 & 4.06 + 126.28 & 0.08 | 8.38 & 1.18 & 1.34  \\\cline{2-10}
\multicolumn{1}{|l|}{} & \ours\   & 99.08\% & 10140 & 0& 0.142 & 4.05 & 1.11  & 0.25 & 0.29 \\ \hline\hline

\multicolumn{1}{|l|}{\multirow{7}{*}{HIV-real}} & Virus-VG & 83.36\% & 8637 & 0 & 3.384 & 3550 + 440.71 & 26.85 | 0.80& - & - \\ \cline{2-10}
\multicolumn{1}{|l|}{} & VG-Flow & 89.99\% & 5950 & 0 & 1.100 & 3550 + 1499.61 & 26.85 | 1.47& - & -\\ \cline{2-10} 
\multicolumn{1}{|l|}{} & viaDBG  & 87.25\% & 1813& 0& 0.197 & 17.24  & 3.74 & - & -  \\ \cline{2-10} 

\multicolumn{1}{|l|}{} & PEHaplo & 91.43\% & 1262 & 0 & 0.074 & 68.34 & 3.48 & - & -  \\\cline{2-10}
\multicolumn{1}{|l|}{} & PredictHaplo & 90.21\% & 8702 & 0 & 0.287 & 4.71 + 100.75 & 0.09 | 0.87 & - & -  \\\cline{2-10}
\multicolumn{1}{|l|}{} & CliqueSNV & 72.17\% & 8676  & 0 & 1.125 & 4.71 + 136.68 & 0.09 | 9.03 & - & -  \\\cline{2-10}
\multicolumn{1}{|l|}{} & \ours\   & 90.85\% & 2267 & 0& 0.292 & 3.73 & 1.07 & - & -  \\ \hline
\end{tabular}
\end{table*}

In this section, we present several evaluations to verify the quality of the obtained contigs. 

The \textit{de novo} reconstruction methods, Virus-VG, VG-Flow, viaDBG, and PEHaplo, and the reference-based methods, PredictHaplo and CliqueSNV, were configured with the settings given by their authors \cite{virusvg,vgflow,viaDBG,pehaplo,predicthaplo,cliquesnv}. We use the optimal values to determine the solid \kmers\ for viaDBG and ViQUF. A discussion on the impact of these values can be found in the supplementary material.

The results of the reference-based methods show high variability in terms of genome fraction and error rate; however, N50 is almost the length of the entire haplotype for all datasets. For example, on HCV-10, using the same alignment file, CliqueSNV only retrieves one genome, obtaining 9.97\% of genome fraction, whereas PredictHaplo retrieves nine out of 10 genomes. The opposite situation occurs in POLIO-6 and ZIKV-15. On HIV-real, PredictHaplo performs well in all metrics, whereas CliqueSNV retrieves less than four of five genomes, 72.17\% with a higher error rate.


The results of the \textit{de novo} approaches show that although all tools have an overall good performance in terms of genome fraction, on average ViQUF achieved the highest level of genome fraction retrieved. However, there are some cases where large differences appear. The first case is POLIO-6, where  viaDBG  with the optimal configuration retrieves only 73\% of the genome fraction, whereas the remaining methods range from 89\% to 99\%. In ZIKV-15, VG-Flow achieves a genome fraction of 83\%, whereas the rest cover more than 89\% of the genome. This difference may seem small, but the loss of one complete haplotype causes a 6\% loss in genome fraction for 15 haplotypes. Finally, for HIV-5, PEHaplo achieves only 78\% of the genome fraction with twice the error rate  compared to the remaining \textit{de novo} tools.

Analyzing the  length of the obtained contigs, N50 shows that viaDBG loses by far against the rest of the tools. However, ViQUF and PEHaplo are slightly below Virus-VG and VG-Flow, but they can obtain competitive results.
We note that DBG methods usually produce more fragmented assemblies than overlap methods because of using $k$-mers instead of full-length reads. 

We also evaluated the number of misassemblies and the error rate. All tools perform well in these aspects. There are only a few examples where VG-Flow, Virus-VG, and PEHaplo have quite high error rates. Virus-VG and VG-Flow have an extremely good performance on N50 but with many errors in HIV-real. Therefore, we assume that the large N50 is due to the mixing of multiple haplotypes. A similar situation occurs for PEHaplo in ZIKV-15 where the error level is very high, but N50 and genome fraction are almost perfect.

The previously published values \cite{virusvg,vgflow} did not show this anomaly. The reason is probably that Virus-VG and VG-Flow were evaluated with QUAST version 4.3, whereas we used MetaQUAST from QUAST version 5.0. Furthermore, in previous evaluations of Virus-VG and VG-Flow, contigs were allowed to align to several haplotypes, whereas we only allowed one contig to align at one place.
Although one can think that this evaluation method may impact the genome fraction level, making it smaller, this is not the case since our experiment shows similar values.

Another interesting result is the misassemblies for HIV-5, where all \textit{de novo} tools have at least two misassemblies. 
The reason for these strange results is the terminal repeats of the HIV genome.
Therefore, no matter the graph methodology, if the shared region between the start and end of the genome is longer than the $k$-mer, for DBG methods, or the average read, for overlap-based methods, they will probably extend longer  than  the actual genome. 

The overall comparison between reference-based and \textit{de novo} methods supports \textit{de novo} approaches as the most reliable ones. However, the divergence levels in these datasets range from 1\% to 12\%, which might be high for a short-term virus infection. We have performed two extra experiments in Section \ref{boundaries} by simulating two new datasets with a divergence between 0.1\% to 2\%. 

\subsection{Haplotype abundance estimation}

This section evaluates the precision of the haplotype relative frequency estimations. PEHaplo and viaDBG are excluded since they do not provide this information, as stated before.
Again, we used the last version of MetaQUAST with the ``--unique-mapping'' set. If it is not set, one contig might be aligned to several haplotypes; thus, misplacing it and making frequency estimation harder or even impossible.

To evaluate the haplotype relative frequency estimation error, we use two different measures. More specifically, the Mean Estimation Error (MEE), which is the average error per haplotype, and the estimation error standard quasideviation ($\hat{S}_{EE}$), which measures the amount of dispersion in the frequency error estimations.

We will only consider the set of contigs with N50 larger than 75\% of the genome length; thus, the contigs assigned to each haplotype should be long enough to estimate the haplotype frequency. Additionally, more than one contig could cover a particular haplotype. This can happen in  complex or simple datasets at the beginning and end of the haplotype, where frequencies are expected to be less reliable. For a fair comparison, the estimated frequency for one haplotype belongs to the longest contig in these situations. 
Besides, some short contigs sometimes appear in assemblies because of repetitive sections or failures in the assembly. These contigs are expected to have lower or inaccurate frequencies; thus, they will not be considered in the evaluation.

Table \ref{table3} presents the overall results of the five tools on the different datasets. In this experiment, from the \textit{de novo} perspective, there is no clear winner. Virus-VG and VG-Flow perform better than ViQUF for HCV-10 and POLIO-6. However, ViQUF outperforms these techniques for ZIKV-15 and HIV-5, obtaining lower estimation errors and less variability. Overall, the three tools can properly estimate the haplotype frequencies for the datasets. These estimations are accurate for HCV-10 and ZIKV-15, with estimation errors per haplotype below 2\% on average. However, the three techniques obtain lower levels of accuracy for POLIO-6 and HIV-5. One possible explanation is the high level of similarity between haplotypes. This high level of similarity (Table \ref{table1}) means that the paired-end information of unitigs shared between two or more haplotypes point to other sets of unitigs that are also shared. Therefore, this complicates the estimation of the abundance frequencies. Nevertheless, all tools can assemble haplotypes with high levels of accuracy.
However, for POLIO-6, the differences in abundance 
create even more problems on low abundance haplotypes with lower levels of paired-end information and a lower level of reliability on their frequencies. Finally, the reference-based methods achieve  worse results, especially when they do not retrieve all haplotypes. The supplementary material contains individual estimations per haplotype and dataset for further details about which haplotypes have been retrieved and the error in each estimation.

\subsection{Time performance and memory consumption}

 We measured the running time and peak memory usage required by  the seven tools. For evaluation, we used only one core to execute the different techniques. For VG-Flow, Virus-VG, PredictHaplo and CliqueSNV, elapsed time and memory consumption are split into two terms separated by ``+'' and ``|'' symbols, respectively. For VG-Flow and Virus-VG, the first time term refers to the time required to build the preassembled contigs that the tools need, while the second term is the actual time for running the tool. The reference-based methods, PredictHaplo and CliqueSNV, require an alignment file; thus, the first term refers to the time employed to perform the alignment process. We used SAVAGE \cite{savage} to obtain the preassembled contigs as suggested by the authors of VG-Flow and Virus-VG. Additionally, we used BWA \cite{bwa} to obtain the alignment files for reference-based methods.
For running time, we used the symbol ``+'' to report the results since the entire running time is the sum of the time required by SAVAGE or BWA and the one required by a specific assembly tool. 
We use the symbol ``|'' for memory consumption, since peak memory consumption of the entire process is the maximum of both results. We decided to show the information separately because one could use another multiassembly, such as viaDBG or ViQUF, or alignment tool, such as Bowtie or Bowtie2, for this first step, which might be beneficial in terms of running times or memory consumption.


 First, running times for \textit{de novo} in Table \ref{table3} show that ViQUF is the fastest approach, faster than viaDBG, which is also a DBG-based method, and PEHaplo, which (although it is an overlap-based method) can correctly reduce the input data several orders of magnitude. Finally, it is significantly faster than Virus-VG and VG-Flow, based on overlap graphs.  ViQUF is between 275 and 1070 times faster than SAVAGE, followed by Virus-VG, and between 275 and 1354 times than SAVAGE, followed by VG-Flow. 

These results were expected since the DBG-based methods skip reads alignment, which takes quadratic time. Furthermore, ViQUF and viaDBG are faster because VG-Flow and Virus-VG require aligning reads to contigs (which takes $O(nm)$, where $m$ is the number of contigs and $n$ is the number of reads) even without considering SAVAGE execution time, thus removing reads alignment. 
Compared with viaDBG and PEHaplo, ViQUF is between 5 and 80 times faster. The current implementation of ViQUF is not parallelized. The most time-consuming part of ViQUF is building APAG, which could, in principle, be computed in parallel for each pair of adjacent nodes. However, the practical impact of parallelization would be negligible since even without parallelization, ViQUF finishes in less than 5 minutes on all datasets. 

For the reference-based methods, running times include the time required for alignment, even though it is negligible compared with the actual assembly time. Running times for reference-based methods are larger than those for the DBG-based methods; specifically two to four times slower than viaDBG and 10 to 40 times slower than ViQUF. These are significantly smaller than the times of Virus-VG and VG-Flow, and they are comparable with those of PEHaplo.
 
 Finally, our entire approach is more efficient in terms of memory consumption since SAVAGE+Virus-VG and SAVAGE+Virus-Flow require 12 to 25 times the space required by ViQUF, and ViQUF requires less than half of the memory required by viaDBG and PEHaplo. However, the two reference-based methods show different behavior. While PredictHaplo is comparable to ViQUF, even outperforming its results for HIV-real, CliqueSNV requires from eight to 10 times more memory than PredictHaplo and ViQUF.

\subsection{Testing {\em de novo} approaches boundaries}\label{boundaries}

The experiments shown so far reveal an advantage of {\em de novo} methods against the reference-based ones. However, all datasets  used in those experiments have relatively high divergence ratio between the haplotypes. Therefore, the previous sections do not include a scenario where the viruses have not evolved a lot from the common ancestor.

To test this scenario and, at the same time, test the {\em de novo} algorithms boundaries, we used FAVITES \cite{bty921} and DWGSIM to simulate
two different datasets with five haplotypes with low divergence ratios, from 0.1\% to 2.0\%. Furthermore, we have established two different abundance levels to test two different situations. In the first case we have the same abundance, 20\%, for each simulated haplotype
and a coverage of 20000$\times$. In the second case, we have a more realistic situation with different abundances, 33.3\%, 27.6\%, 20\%, 12.5\%,5\%, and 2.5\% but with a very high coverage of around 40000$\times$. Through these experiments, our goal is to test the limits of {\em de novo} and reference-based approaches in terms of sensitivity and scalability.

Table \ref{ref:table_supp_extra} summarizes the results of these experiments. In the first dataset, where abundance for each strain is the same, the reference-based methods, PredictHaplo and CliqueSNV, slightly outperform the {\em de novo} approaches. The percentage of genome retrieved is almost the same, except for CliqueSNV, which is able to obtain the five entire haplotypes with almost no error. However, mismatches here are really important, for example,  PEHaplo's 0.8 might be really problematic when maximum difference between strains is around 2\%, in the case of ViQUF this number is smaller, 0.314, but it is still higher than that of the reference-based methods. In the second dataset, both reference and {\em de novo} based methods retrieve the entire set of haplotypes with a N50 as long as the actual genomes. However, in this case, there is no advantage of using reference-based methods because ViQUF retrieves the same percentage of the genome with no errors, as PredictHaplo. For the second dataset, we were not able to get results for SAVAGE+VG-Flow and CliqueSNV. SAVAGE could not produce long enough contigs in any of the experiments, whereas CliqueSNV requires more than 64Gb and thus, we could not run it on the same machine we used for the rest of the experiments. Table \ref{ref:table_supp_extra} also shows the time needed to run all the experiments. While the first dataset has the same coverage as the ones used in previous sections, the second has twice the coverage. This second dataset was simulated in this way to be able to measure the scalability of the approaches.  As the results show, with the exception of CliqueSNV, the methods are scalable. PredictHaplo has the best scalability ratio, whereas ViQUF is the one with the lowest absolute runtimes.

Overall, the results exposed in Table \ref{ref:table_supp_extra} show that reference-based methods are slightly more reliable than \textit{de novo} approaches when the divergence between the strains is very low, specially CliqueSNV, which has an astonishing performance. However, \textit{de novo} approaches are on average faster and use less memory than the reference-based methods, even if we do not take into consideration the required time for the alignment.

\begin{table*}[]
\centering
\footnotesize
\caption{Results for the viral quasispecies assembly tools, de novo and reference-based, for small divergence ratios and extremely high sequencing coverage. For reference-based methods, the elapsed time column shows the time required for alignment plus the tool execution time. }
\begin{tabular}{l|l|r|r|c|c|r|}
\cline{2-7}
                                                            &              & \% Genome & N50 & Misassemblies & Mismatches & Time elapsed (mins) \\ \hline
\multicolumn{1}{|l|}{\multirow{5}{*}{Same abundance }}     & ViQUF        &  83.97\%         & 9975    &  0     &      0.314      &        4.53   \\ \cline{2-7} 
\multicolumn{1}{|l|}{}    & SAVAGE+VG-Flow      &     *      &  *    &     *          &    *       &        *                                             \\ \cline{2-7} 
\multicolumn{1}{|l|}{}                                      & PEHaplo      & 80.00\%   &  10000   &  0  &    0.807    &     227.96    \\ \cline{2-7} 
\multicolumn{1}{|l|}{- 20000$\times$ coverage}                                      & PredictHaplo &  80.00\%         &  10000  & 0  & 0.205           &     10.53 + 442.78         \\ \cline{2-7} 
\multicolumn{1}{|l|}{}                                      & CliqueSNV    &  100.00\%   &  10000   &     0      &    0.100        &   10.53 + 34.93             \\ \hline
\multicolumn{1}{|l|}{\multirow{5}{*}{Different abundances }}   & ViQUF        &  99.94\%         & 9995 &      0     & 0.000           &       25.56  \\ \cline{2-7} 
\multicolumn{1}{|l|}{} & SAVAGE+VG-Flow      &    *       & *    & * & * & *                                         \\ \cline{2-7} 
\multicolumn{1}{|l|}{}                                      & PEHaplo      &  100.00\%         & 10000     & 0    & 0.604    &       1051.56       \\ \cline{2-7} 
\multicolumn{1}{|l|}{- 50000$\times$ coverage}                                      & PredictHaplo &    100.00\%       & 10000    &  0 & 0.100 & 23.73 + 1475.23             \\ \cline{2-7} 
\multicolumn{1}{|l|}{}                                      & CliqueSNV    &    *       &  *   &    * &     *       &       *       \\ \hline
\end{tabular}
\label{ref:table_supp_extra}
\end{table*}

\section{Conclusions}

In this work, we proposed \ours\ to {\em de novo} assemble viral quasispecies. The methodology of \ours\ bears some similarity to our previous work, viaDBG \cite{viaDBG}, which also uses paired-end information to split nodes of a DBG into haplotypes. However, \ours\ includes several important improvements. First, we present a mathematically rigorous way of determining solid \kmers\ using kernel density estimation, resulting in better estimates for the abundance threshold of solid $k$-mers as shown by our experiments. Second, we use path cover to split the adjacent nodes into haplotypes, whereas viaDBG finds cliques in the reachability graph of paired-end nodes. Third, \ours\ uses path cover to find haplotypes and their abundances; thus, considering the coverage in this stage. However, viaDBG only reports nonbranching paths as contigs. Our experiments show that these improvements allow us to build longer contigs with comparable accuracy fast.


In the overall results, all methods show remarkable performance. 
In comparison, Virus-VG, VG-Flow,and the two reference-based methods (PredictHaplo and CliqueSNV) perform slightly better in average contig length than their counterparts. However, the error rate is higher than expected for some cases, and the reference-based methods lose some haplotypes. This is probably because of the high divergence between the haplotypes in each sample. 
However, viaDBG and ViQUF exhibit a more robust behavior without an unexpectedly high error rate for any dataset. Furthermore, ViQUF shows  high levels of sensitivity, retrieving more than 90\% of the genome fraction.
ViQUF outperforms all methods when comparing memory consumption and runtime, especially VG-Flow and Virus-VG. It is also remarkable that PredictHaplo exhibits a great performance in running time and memory usage, especially in memory usage, where it is comparable with ViQUF and even better for real data. For this dataset, PEHaplo achieves the best performance in terms of genome fraction and error rate.
ViQUF is also faster than viaDBG, using noticeably less memory. Moreover, 
ViQUF can provide haplotype abundance, whereas viaDBG cannot. 


In summary, ViQUF obtains long accurate contigs consuming fewer resources than other methods. 
 Furthermore, ViQUF can produce competitive haplotype frequency estimations compared with current state-of-the-art tools like Virus-VG and VG-Flow. 
However, our experiments were mostly conducted on synthetic data.
Therefore, it is highly desirable to obtain new real datasets to conduct more thorough experimental evaluations in our future studies.

As a comparative conclusion, the viral quasispecies assembly and quantification have several good but different solutions. Some solutions are better than others depending on constraints, such as runtime, memory resources, accuracy required, and properties of a specific dataset. For example, when the sequencing depth is sufficiently high and uniform over each haplotype, and all haplotypes are completely present in the sample, ViQUF or viaDBG can be the best solution when running time and memory consumption are severe constraints. However, under the same conditions but without constraints on running time and memory, using a smart grid of parameters with SAVAGE+VG-Flow could lead to the most accurate results. Furthermore, if the average similarity between haplotypes is below 1\%, the most accurate and consistent results are provided by reference-based methods. 

\section*{Acknowledgments}

This work has received funding from the EU H2020  under the Marie Sklodowska-Curie [GA 690941]. We wish to acknowledge the support received from the Centro de Investigación de Galicia ``CITIC'', funded by Xunta de Galicia and the European Union (European Regional Development Fund- Galicia 2014-2020 Program), by grant ED431G 2019/01. This work was also supported by Xunta de Galicia/FEDER-UE under Grants [ED431C 2021/53; IG240.2020.1.185; IN852A 2018/14]; Ministerio de Ciencia e Innovación  under Grants [TIN2016-78011-C4-1-R;  FPU17/02742; PID2019-105221RB-C41; PID2020-114635RB-I00]; and the Academy of Finland [grants 308030 and 323233 (LS)]. The authors also thank David Posada for his advice on viral evolution.

%
%
%
 
\bibliographystyle{IEEEtran}
\bibliography{main.bib}

\end{document}


\title{ Supplementary Material }

\maketitle

\section*{Contents}
\makeatletter\@starttoc{toc}\makeatother

\section{Background}\label{sec:background}
In this section, we introduce some basic concepts used by our method, such as the network flow problem and kernel density estimation.

\subsection{Min-cost network flow problem}

A flow network is a tuple $N = (G, b, q)$, where $G = (V,E)$ is a directed graph, $b$ is a function assigning a \textit{capacity} $b_{uv}$ to every arc $(u,v) \in E$, and $q$ is a function assigning an exogenous flow $q_v \in \mathbb{Z}$ to every node $v \in V$, such that $\sum_{v\in V}q_v = 0$. Then, we have a flow over the network $N$, if for every arc $(u,v) \in E$ the flow $x_{uv} \in \mathbb{N}$ satisfies two conditions:
\begin{enumerate}
    \item $0 \leq x_{uv} \leq b_{uv}$ for every $(u,v) \in E$
    \item $\sum_{u\in V}x_{vu} - \sum_{u\in V}x_{uv} = q_v$, for every $v \in V$
\end{enumerate}

In the min-cost flow problems, like in many other cost-flow optimization problems, a cost function $c_{uv}(x)$ is given for every $(u,v) \in E$, and the objective is to find the flow which minimizes:
\begin{equation*}
    \sum_{(u,v) \in E}c_{uv}(x_{uv}).
\end{equation*}

This problem can be solved in polynomial time when $c_{uv}(x)$ is a convex cost function. One of the most common cost functions is $c_{uv}(x) = x ^2$. 

\cite{traph} proposed a min-cost flow method for estimating transcript expression with RNA-Seq. We will use a similar approach as part of our method.

\subsection{Kernel density estimation}
Kernel density estimation is a very powerful mathematical tool for estimating the probability density function of a random variable. In practice, it creates a smooth curve given a set of data. More concretely, it can be expressed as:
\begin{equation*}
    \hat{f}_h(x) = \frac{1}{n}\sum_{i = 1}^{n}K_h(x - x_i)
\end{equation*}

\begin{equation*}
    K_h(y) = \frac{K(y)}{h}
\end{equation*}
where $x_i$ are the sample data points, $n$ is the sample size, $h$ is the window size or the bandwidth, 
and $K(\cdot)$ 
is a kernel function, such as Gaussian, uniform,  or Epanechnikov, among others \cite{Jones90}. Once $\hat{f}_h$ is estimated, \leena{the minima and maxima of $\hat{f}_h$ are obtained by computing the zero crosses of its derivative.}

\section{Example of paired unitigs}
\tolerance 500 \pretolerance 500

\begin{figure}[t]
	\centering
	\includegraphics[width=0.45\textwidth]{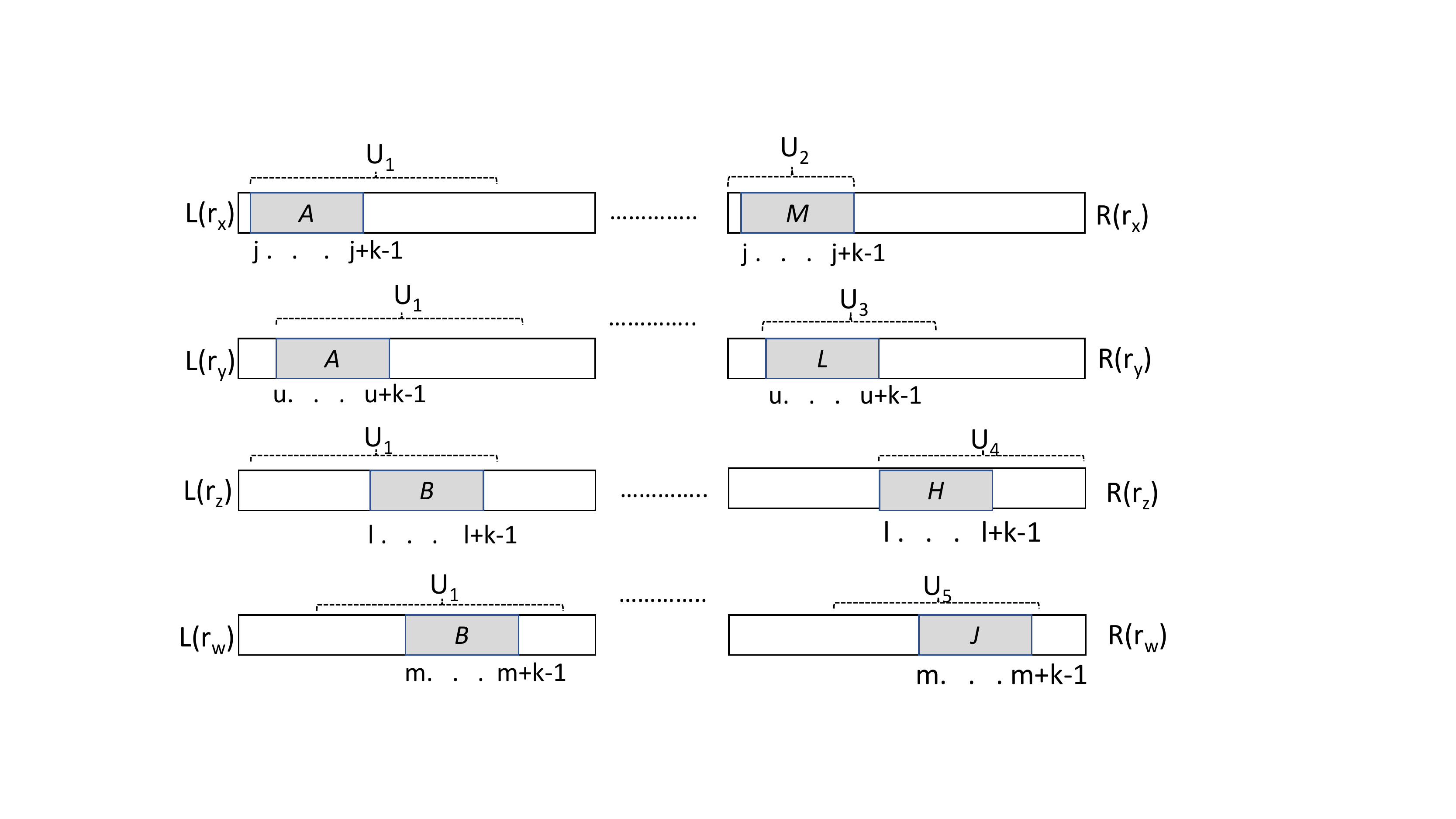}
	\caption{Extracting paired unitigs from paired-end reads $ P(U_1)=\{U_2,U_3, U_4,U_5\}$.}\label{fig:pairedEnd}
\end{figure}

In Figure \ref{fig:pairedEnd},  the unitig $U_1$ contains (among others) the \kmers\ $A$ and $B$. Observe that $L(r_x)$ contains the \kmer\ $A$   at positions $j\ldots j+k-1$, and in those positions of $R(r_x)$ we find the \kmer\ $M$. In $r_y$, $A$ is at positions $u\ldots u+k-1$ of $L(r_y)$, whereas positions  $u\ldots u+k-1$ of $R(r_y)$ contain the \kmer\ $L$. Therefore, $P(A)=\{M, L\}$. From the reads $r_z$ and $r_w$, we obtain that $P(B)=\{H, J\}$.  Since $M$ is located inside $U_2$, that is ${\mathcal U}(M) =U_2$, and ${\mathcal U}(L)=U_3$, ${\mathcal U}(H)=U_4$, and ${\mathcal U}(J)=U_5$, then finally $P(U_1)=\{U_2, U_3, U_4, U_5\}$. 

\section{Coverage of an edge of the DAG}

\leena{To split the nodes of the assembly graph into haplotypes, we build $DAG_{ij}$ for each pair of adjacent nodes $U_i$ and $U_j$ in the assembly graph. The nodes of $DAG_{ij}$ are the pair of adjacent nodes and the nodes pointed by their paired-end information. We add an edge between two nodes if there exists a path between the corresponding nodes in the assembly graph.}
Here, we present the method to compute the coverage of an edge $(s,e)$ of a $DAG_{ij}$. 

\subsection{Initial coverage estimation}

\leena{$DAG_{ij}$ is constructed to analyze haplotypes that pass through $U_i$ and $U_j$. Therefore when computing the coverage of an edge $(s,e)$ in $DAG_{ij}$, we should only consider haplotypes that pass through all nodes $U_i$, $U_j$, $s$, and $e$. {\em The coverage of an edge $(s,e)$ in $DAG_{ij}$} is thus defined as the total coverage of all haplotypes that pass through the nodes $U_i$, $U_j$, $s$, and $e$.} For this, we analyze the coverages of the paths connecting $U_i$, $U_j$, 
$s$, and $e$ \textit{in the AG}. The coverage of an edge of the AG is the number of reads supporting that edge.

    \begin{figure}[t]
    \centering
           \includegraphics[scale=0.6]{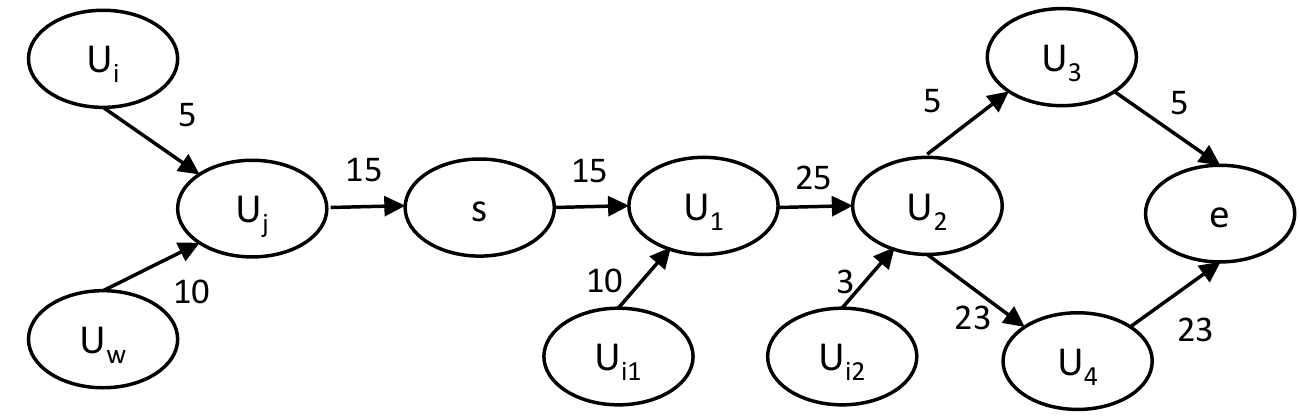}
\caption{Extract of an AG used to compute the coverage of the edge between $s$ and $e$ in a DAG $DAG_{ij}$.}   \label{fig:flow_t}    
\end{figure}

      To illustrate the process, in Figure \ref{fig:flow_t}, we can see an  extract of an  AG.  Observe in the figure   
      that the coverage of the edge $(s,U_1)$ is 15, but that coverage cannot be used as that of $cov(s,e)$ in the $DAG$, since it includes 10 reads that are actually reads corresponding to the  haplotype  that passes through $U_w$ and $U_j$ \leena{but not through $U_i$}, and therefore the coverage corresponding to that haplotype must not be included in the coverage of $(s,e)$ in $DAG_{ij}$, which is computed for detecting the haplotypes that pass through $U_i$ and $U_j$. 
      The mixture of coverages coming from different haplotypes in the paths  from $s$ to $e$ makes it impossible to use simple estimators such as maximum, minimum, or mean \leena{of the coverages of edges on the path from $s$ to $e$}.
      
    
     Given a node $U_k$ of the $AG$, let $N^+_{U_k}$ be the list of out-neighbors of $U_k$, $N^-_{U_k}$ be the list of in-neighbors, and $cN^+_{U_k}$ and $cN^-_{U_k}$ the \leena{lists of the coverages of the corresponding edges}.
        For example, in our example of Figure \ref{fig:flow_t}, $N^+_{U_2}=\{U_{3},U_{4}\}$ and $N^-_{U_2}=\{U_1, U_{i2}\}$, and $cN^+_{U_2}=\{23,5\}$ and $cN^-_{U_2}=\{25,3\}$.
   
    We are going to traverse the $AG$  through all paths connecting $U_i$, $U_j$, $s$, and $e$, analyzing $cN^+$ and $cN^-$ in each node, in order to derive a reliable coverage. 
        During a traversal of the $AG$ that has already processed the path $p=\{U_i, U_j, \ldots, s, U_1, \ldots, U_k\}$,  roughly, we maintain a bag $B_k$ containing the   coverages of edges reaching  nodes of $p$ and coming from nodes that are not in $p$, thus they correspond to haplotypes different from the one we are following. 
    Therefore, when $e$ is reached, the coverage of that path is the coverage of the edge reaching $e$ minus the coverages in the bag. 
        However, on our way towards $e$, we will find outgoing edges leading to paths that do not reach $e$, so the coverage of those edges should be removed from the bag.
    
    Therefore, when we process a new node with one or more outgoing edges leading to a node which is not on our way to $e$, we need a method to distribute the incoming coverages stored in the bag among those outgoing nodes. A first naive idea is to search for  coverages in the bag matching the coverages of outgoing edges corresponding to paths that do not reach $e$. We remove those values from the bag,   assuming that those values correspond to haplotypes that entered in $p$ in a previous node and now, we found a node where those haplotypes are leaving the path we are following towards $e$. The problem with this approach is that each incoming edge may correspond to several haplotypes, and therefore they might leave our path $p$ in different points, and thus this approach  is not capable of dealing with this situation. 
    
    Therefore, we need a more complex method. As explained, the method is based on traversing the $AG$ from $U_i$ until $e$ through all possible paths, carrying a bag with the incoming coverages.  When a traversal finds a fork, it  distributes those coverages  among the nodes in the fork, by using the following iterative minimization problem.  
    \begin{itemize}
     
        \item Let $p_k=\{U_i, \ldots, U_k\}$ be path of the $AG$ followed so far.
           \item  Let $cov(p_k)$ the coverage estimation of the path $p_k$.
        
        \item Let $iC = \{ic_1,...,ic_n, cov(p_k)\}$ be  the coverages of $n$ incoming edges to $p_k$ with the coverage of $p_k$ included as well.
        \item Let $oC = \{oc_1,...,oc_b\}$ be the coverages of the $b$    outgoing edges from $U_k$.
    \end{itemize}
    \leena{The task is then to assign the incoming coverages to the outgoing coverages so that they correspond to each other as well as possible while simultaneously avoiding to split the incoming coverages, i.e. assigning them partially to several outgoing coverages.}
    
    \leena{For each $v\in \{1...n\}$ and $u\in \{1...b\}$, we define a variable \borja{$x_{vu} \in \{0,1\}$} which indicates \jose{whether the} incoming coverage $ic_v$  \jose{is assigned to the} outgoing coverage $oc_u$. In the first iteration,  \jose{$iC=B_{k} \cup cov(p_k)$, \borja{where $B_{k}$} is the bag of incoming coverages on path $p_k$ and $cov(p_k)$ the coverage estimate of the traversed path, } and $oC=cN^+_{U_k}$. In a single iteration, we assign one incoming coverage completely to an outgoing coverage, whereas several incoming coverages can be assigned to the same outgoing coverage. 
    \borja{Because $cov(p_k)$ has been added to the set of incoming coverages, in each iteration it can only be assigned to one outgoing edge. However, we allow the coverage of the path $cov(p_k)$ to be split by including the remaining of it in successive iterations}}.
    \leena{Furthermore, we want to minimize the difference between the assigned incoming coverages and the outgoing coverage of each edge. More formally, the minimization problem is defined as follows:
     \begin{equation}
    \begin{aligned}
        \mathrm{min} \quad \sum_{u = 1}^b \left | oc_u - \sum_{v = 1}^n(ic_v x_{vu}) \right | \\
        \textrm{subject to:} \quad \sum_{u=1}^bx_{vu} = 1,\forall v \in \{1...n\} \\
        x_{vu} \in \{0,1\}, \forall vu \in \{1...n\}\times\{1...b\} \\
    \end{aligned}
    \label{equation1}
    \end{equation}
    The constraints make sure that one incoming coverage is assigned completely to a single outgoing coverage.}
    
    
    \leena{Let iteration $i$ be defined by $oC_i$, the list of outgoing coverages still not completely used by the assigned incoming coverages, and $iC_i$, the remaining incoming coverages available. Then iteration $i+1$ is defined as follows. We subtract from each outgoing coverage the coverage of the incoming edges assigned to that outgoing edge:
    \begin{equation}
    oC_{i+1} = \{\max(oc_u - \sum_{v = 1}^n ic_v x_{vu},0) \quad \forall u \in \{1...b\}\}
    \end{equation}
    Note that if the sum of assigned incoming coverages exceeds the outgoing coverage of the edge, we set the outgoing coverage to zero for the next iteration. Similarly, we will subtract from the incoming coverages the coverage of the edge they have been assigned to. However, more than one incoming coverage can be assigned to the same outgoing coverage. If the sum of the assigned incoming coverages exceeds the outgoing coverage,  we will decrease the incoming coverages starting from the largest incoming coverage. Assuming that $ic_v$ are sorted in ascending order, we thus get:
    \begin{equation}
        \begin{aligned}
    iC_{i+1} = \{ \max (ic_v-\max(oc_u-\sum_{v'=v+1}^n x_{v'u} ic_{v'}, 0), 0)\} \\
     \textrm{where } u \textrm{ is such that } x_{vu}=1 \quad \forall v \in \{1...n\}\}
    \end{aligned}
    \end{equation}
}
    

     
     \leena{The iterative process ends when either all incoming coverages have been assigned to outgoing coverages or all outgoing coverages have been totally used.}     
When the iterative process ends, the new estimation of the coverage of the new extended path ($cov(p_{k+1})$)  is obtained  by  subtracting from the coverage of the assembly graph edge $(U_{k}, U_{k+1})$ the sum of incoming coverages in $B_k$ assigned to the branch of $U_{k+1}$. \jose{These assigned coverages form the new bag $B_{k+1}$}. \borja{As a reminder, $B_{k+1}$ does not contain the fraction of $cov(p_k)$ assigned to the branch; \susana {notice that the inclusion of $cov(p_k)$} is only for minimization problem purposes.}
     
     This approach not only allows the incoming coverage to be distributed among several outgoing edges, but also allows estimating the coverage of the haplotype. In this way it is possible to avoid the traversal of  paths without assigned coverage, speeding up the build process of the DAGs and leading to less tangled DAGs.

    \bigskip
    \noindent
    \leena{There can be several paths passing through $U_i$, $U_j$, $s$, and $e$ and often these paths share a prefix. Next we will explain how to traverse all these paths to compute the coverages.}
          The process is as follows: let $\mathcal{T}$ be a list of traversals of the $AG$. Each traversal is  a path of the $AG$ that starts at $U_i$, passes through $U_j$,  and contains the nodes processed so far on its way through $s$ towards $e$. 
      
      Initially $\mathcal{T}$ contains  one traversal  $t_j$,  its path only contains \leena{$U_iU_j$}, and its coverage ($cov(t_j)$) is the coverage of the edge $(U_i,U_j)$.  Each traversal has a bag, initially, the bag of  $t_j$ is \leena{$B_{j}=cN^-_{U_j}$}.  
      
      \leena{Then we will repeatedly remove one traversal from $\mathcal{T}$ until it is empty, compute the  extensions of the traversal, and insert them into $\mathcal{T}$ unless they have reached $e$, their coverage has dropped to zero, or they are not supported by the paired-end information\footnote{That is, the new node to be added  should be in the paired-end information of at least one node in the path traversed so far.}. The extensions of a traversal $t_\lambda$ are computed as follows. Suppose $p_\lambda=\{U_i,...U_{k}\}$ is the path followed by traversal $t_\lambda$ so far and $B_k$ is the bag of incoming coverages associated with the traversal. We create a new traversal for each node $U_{k+1}\in N^+_{k}$ by concatenating $p_\lambda$ with $U_{k+1}$. If $U_k$ has more than one outgoing edge, we use the minimization problem above to distribute the incoming coverages $B_k$ appropriately and to compute the incoming coverages for the extended traversal and the coverage of the extended traversal. If $U_{k+1}$ has more than one incoming edge, the coverages in $cN^-_{U_{k+1}}$ except for the coverage of the edge $(U_k,U_{k+1})$ are added to $B_k$ to form $B_{k+1}$. If the coverage of the extended traversal drops to zero or it contradicts the paired-end information of $p_\lambda$, we stop the traversal and do not add it to $\mathcal{T}$. Finally, the coverage of the edge $(s,e)$ in $DAG_{ij}$ is the sum of coverages of all traversals that reached $e$.
      }

To illustrate the process,  we are going to compute the coverage of the edge connecting $s$ and $e$ in the $DAG$ $DAG_{ij}$ for the AG shown in Figure \ref{fig:flow_t}.

 Table \ref{tab:trace} shows  in row  number 1 the initial state of $\mathcal{T}$ with only one traversal $t_{ij}$, its bag $B_j = cN^-_{U_j}=\{10\}$, and the initial coverage estimation ($cov(t_{ij})$) is set to $cov(U_i,U_j)$.
 
 Row 2 processes $s$, which since it only has one incoming and one outgoing edge does not produce any change.
 Row 3 processes $U_1$, that, in addition to the edge coming from $s$, has also an incoming edge coming from $U_{i1}$, and then,  the coverage (10) of that edge is added to the bag $B_{1}$.
 
 Row 4 processes the out neighbors of $U_1$, this adds only $U_2$. Then,  the coverage of the  edge coming from $U_{i2}$ is added to the bag, that is $B_{2}=\{10,10,3\}$.

\begin{table}[tbh]
\centering
\scriptsize
\begin{tabular}{|l|l|l|l|l|}
\hline
\multicolumn{1}{|c|}{$\#$}  & \multicolumn{1}{c|}{$\mathcal{T}$} & \multicolumn{1}{c|}{$U_{k}$}                                                           & \multicolumn{1}{c|}{$B_{k+1}$}                                            &  \multicolumn{1}{c|}{$cov$} \\ \hline \hline
1& $t_{ij}$ &$U_i$&$B_{j}=\{10\}$&$5$\\ \hline
2                          &                            $t_{ijs}$                                                                   &  $U_j$   & $B_{s}=\{10\}$                                                          &                         $5$ \\ \hline
3                                                 & $t_{ijs1}$       & $s$                                                                  & $B_{1}=\{10,10\}$                                                     &                         $5$ \\ \hline
4                                               & $t_{ijs12}$      & $U_1$                                                                & $B_{2}=\{10,10,3\}$                                                   &                         $5$ \\ \hline
5                                                 & $t_{ijs123}$     &  $U_2$&  \begin{tabular}[c]{@{}l@{}}$B_{3}=\{\}$\\ $B_{4}=\{10,10,3\}$\end{tabular} &            \begin{tabular}[c]{@{}l@{}}$cov (t_{s123})=5$ \\ $cov (t_{s124})=0$\end{tabular}               \\ \hline
6                                             & $t_{ijs123e}$   &     $U_3$  &                             $B_{e}=\{\}$ &                                     cov($t_{ijs123e})=5$         \\\hline
\end{tabular}
\caption{Trace of the computation of the coverage of the edge $(s,e)$ of $DAG_{ij}$.}\label{tab:trace}
\end{table}

    Row 5 shows the processing of the outgoing neighbors of $U_2$. This extracts $t_{ijs12}$ from $\mathcal{T}$, and since $U_2$ has two outgoing neighbors ($U_3$ and $U_4$), two traversals  ($t_{ijs123} $ and $t_{ijs124}$) are created. 
    
    
     \begin{itemize}
        \item 
        We formulate the minimization problem  of Equation (\ref{equation1}) taking $iC=B_{2} \cup \{cov(t_{ijs12}]\}=\{10,10,3,5\}$ and $oC=cN^+_{U_{2}}=\{23,5\}$
            \borja{\begin{equation}
           \begin{aligned}
                \min \quad \left | 23 - 10x_{11} - 3x_{21} - 10x_{31} - 5x_{41} \right | \\ + \left | 5 - 10x_{12} - 3x_{22} - 10x_{32} - 5x_{42}\right |\\
                \textrm{subject to:} \quad x_{11} + x_{12} = 1,\\
                 x_{21} + x_{22} = 1,\\
                 x_{31} + x_{32} = 1,\\
                 x_{41} + x_{42} = 1,\\
                x_{11}, x_{12}, x_{21}, x_{22},x_{31},x_{32}, x_{41},x_{42} \in  \{0,1\}\\
            \end{aligned}
            \label{eq:2}
            \end{equation}}
        \item \borja{An \textbf{optimal possible} solution is $x_{11} = 1$, $x_{21} = 1$,$x_{31} = 1$, $x_{12} = 0$, $x_{22} = 0$, $x_{32} = 0$, $x_{41} = 0$ and $x_{42} = 1$. With those values,  expression (\ref{eq:2}) returns zero.}
        \item That solution assigns all the excess in the bag to the outgoing edge that goes to $U_4$ and thus: $B_{3}=\{10\cdot 0,3 \cdot 0, 10 \cdot 0\}=\{ \emptyset \}$ and $B_{4}=\{10\cdot 1,3 \cdot 1, 10 \cdot 1\}=\{ 10,10,3 \}$.  
        
        \leena{Both the set of incoming and outgoing coverages become empty sets for the next iteration and thus the minimization ends after only one iteration.}
        
        \item Therefore, the coverages of the two traversals are:
        \begin{itemize} 
            \item   $cov (t_{s123})=cov(U_2,U_3)-\sum_{v = 1}^3 ic_v x_{v2}=5-(0+0+0)=5$. Thus $t_{s123}$ is added  to $\mathcal{T}$. 
            \item     $cov (t_{s124})=cov(U_2,U_4)-\sum_{v = 1}^3 ic_v x_{v1}=23-(10+3+10)=0$. Therefore, $t_{s124}$ is not added  to $\mathcal{T}$.
        \end{itemize}
    \end{itemize}
    

    Therefore, in this example, it is clear that all the input excesses found in the path through $U_i,U_j,s,U_1,U_2,U_3$ can be assigned to the path that goes through $U_4$, and thus the path that goes through $U_3$ continues without any excess in its bag, whereas that passing through $U_4$ is discarded.

  Row 6 processes  $t_{ijs123}$ appending the target $e$. Therefore, the coverage of the edge $(U_3,e)$ (5) is the coverage of the traversal passing through $U_3$  ($cov(U_3,e)=5$), since the bag $B_{3}$ is empty. 

    
 Once all traversals reached $e$, we obtain  the coverage of the edge $(s,e)$ in $DAG_{ij}$ as the sum of the coverages of the traversals that reached $e$, in our example, it is only $cov(t_{ijs123e}) = 5$.

\subsection{Coverage readjustment}


The previous section explains how to assign coverages  
to the edges of $DAG_{ij}$. However, there are  some situations where this  procedure  generates unreliable variations of the coverages, building peaks and falls \jose{across the graph}. 
These nodes  will  create \susana{an} incomprehensible exogenous flow that leads to misleading flows. Therefore, they have to be adjusted whenever possible. 

We say that a node $U_r$ of  $DAG_{ij}$ is \textit{out of coverage}  when $\sum cN^-_{U_r} \leq (1+RATIO) \sum cN^+_{U_r}$ or vice versa\borja{, where $RATIO$'s default value is $0.1$ namely a node is \textit{out of coverage} when the incoming or outgoing coverages is 10\% higher than the other}. 

We are going to correct these \textit{out of coverage} nodes in the $DAG_{ij}$.
For this, the DAG is traversed in breadth-first manner making sure that all nodes in $N^-_{U_r}$ have been adjusted before adjusting $U_r$. When a node $U_r$ is classified as \textit{out of coverage}, then a solution is found based on its \textit{in-degree} and \textit{out-degree}. Three situations arise:
\begin{itemize}
    \item \textit{in-degree} $\geq$ 1 and \textit{out-degree} = 1, these cases can be easily solved by setting  $cov(U_r, U_{out}) = \sum cN^-_{U_r}$, where $U_{out}$ is the out neighbor.
    \item \textit{in-degree} = 1 and \textit{out-degree} $>$ 1, in these cases, the coverage provided by the incoming edge must be distributed between the nodes $N^+_{U_r}$ based on how close their coverage is to the $cov(U_i,U_j)$, where $U_i$ and $U_j$ are the nodes of the AG under study, that is, the nodes for which we are computing their $DAG_{ij}$.

    The basic idea is that the closer the coverage of an edge $cov(U_r,U_k)$ $U_k \in N^+_{U_r}$ is to $cov(U_i, U_j)$, the higher coverage will be assigned. 
    Closer coverages are expected to be more likely to be the correct ones. 

    The coverage translation is then:
    \begin{itemize} 
        \item[--] For every node $U_l \in N^+_{U_r}$, we define $\mathit{diff}(U_l) = |cov(U_r,U_l) - cov(U_i, U_j)|$.
        \item[--] The new coverage is
       \end{itemize}
      \end{itemize} 
         $$cov(U_r, U_l) = \borja{cN^-_{U_r}}\frac{|\mathit{diff}(U_l) - \sum_{\forall{U_v \in N^+_{U_r}}}\mathit{diff}(U_v)|}{\sum_{l \in cN^-_{U_r}}\left | \mathit{diff}(U_l)  - \sum_{\forall{U_v \in N^+_{U_r}}}\mathit{diff}(U_v)\right |}$$

\begin{itemize}

    \item \textit{in-degree} $>$ 1 and \textit{out-degree} $>$ 1, in these cases it is necessary to assign the nodes in $N^+_{U_r}$ to each node in $N^-_{U_r}$. To do so, we follow the exact same methodology as we followed to assign the excesses found so far in a traversal of the $AG$ to the outgoing coverages of a given node, when computing the coverage of an edge of the $DAG$. 
\end{itemize}


\section{Experimental evaluation}
\subsection{Threshold computation}

Both viaDBG and ViQUF are based on a de Bruijn graph. This requires to determine the set of \kmers\  used to build that graph. The traditional way to obtain this set  is to calculate a threshold such that the \kmers\ with a higher frequency are selected. 
The value of the threshold  might have a high impact on the results. A method based on the \kmer\ frequency histogram is used by viaDBG, whereas, as explained, ViQUF uses Kernel Density Estimation (KDE). Therefore, our first experiment is designed to check the impact of the  methods used  by ViQUF and viaDBG to determine the threshold.






Table \ref{table2} gathers the thresholds computed for all the datasets. Those results suggest that the new methodology is more aggressive than that of viaDBG, obtaining in general higher thresholds. 
To perform a fair comparison,  viaDBG and ViQUF were ran with their own  threshold  and with the threshold obtained by the other tool. The overall results, included in Table \ref{table3}, show that the viaDBG threshold is quite conservative because the  higher threshold of ViQUF  does not affect the  genome fraction, although an almost negligible grow in mismatches appears. \leena{However, the contiguity of the assembly is in general higher with the higher threshold of ViQUF as seen by the N50 values.}

\begin{table}[h]
\centering
\caption{Threshold values to filter solid \kmers\ obtained for different datasets using an approach based on frequency histogram, as used by viaDBG, or the proposed approach using kernel density estimation (KDE), used by \ours.}\label{table2}
\begin{tabular}{|l|r|r|}
\hline
dataset & Freq. histogram & KDE \\\hline
HCV-10           & 16     & 103   \\
HIV-5             & 21     & 146   \\
POLIO-6          &  12     & 51  \\
ZIKV-15           & 5     & 26   \\
HIV-real          & 128    & 172   \\\hline
\end{tabular}
\end{table}
\begin{table*}[]
\caption{Results for the four viral-quasispecies assembly tools. The table also includes the results for the threshold comparison between viaDBG and ViQUF. \leena{For Virus-VG and VG-Flow, we show the elapsed time and memory usage separated into contig assembly by SAVAGE (first value) and the full haplotype reconstruction (second value).}}\label{table3}
\centering
\footnotesize
\begin{tabular}{l|l|r|r|r|r|r|r|}
\cline{2-8}
&& \multicolumn{1}{c|}{\multirow{2}{*}{\% Genome}} & \multicolumn{1}{c|}{\multirow{2}{*}{N50}} & \multicolumn{1}{c|}{misass-} & \multicolumn{1}{c|}{\% mis-} &\multicolumn{1}{c|}{elap time} & \multicolumn{1}{c|}{memory} \\

dataset& method  & \multicolumn{1}{c|}{}  & \multicolumn{1}{c|}{}& emblies & matches & \multicolumn{1}{c|}{(min)} & \multicolumn{1}{c|}{(GB)}   \\ \hline

\multicolumn{1}{|l|}{\multirow{6}{*}{HCV-10}} & Virus-VG & 99.30\% & 9231   & 0 & 0.002 & 913.48 + 1009.08 & \borja{26.13 | 8.35}  \\ \cline{2-8}   
\multicolumn{1}{|l|}{} & VG-Flow & 99.79\% & 9293   & 0& 0.001&  913.48 + 559.56 & \borja{26.13 | 8.29}  \\ \cline{2-8} 
\multicolumn{1}{|l|}{} & \borjareview{PEHaplo} & 94.78\% & 8661  & 0 & 0.013 & 68.45 & 8.94  \\ \cline{2-8}
\multicolumn{1}{|l|}{} & \borjareview{PredictHaplo} & 89.79\% & 9273 & 0 & 0.044 & 4.11 + 175.73 & 1.14  \\\cline{2-8}
\multicolumn{1}{|l|}{} & \borjareview{CliqueSNV} & 9.97\% & 9273 & 0 & 2.10 & 4.11 + 3494.09 & 17.24   \\\cline{2-8}
\multicolumn{1}{|l|}{} & viaDBG ($t=16$)& 97.72\%  & 8934   & 0   & 0.005&  69.10& 2.81  \\ \cline{2-8} 
\multicolumn{1}{|l|}{} & viaDBG ($t=103$)&  97.18\% & 8936  & 0& 0.010&  68.12 & 2.81  \\ \cline{2-8} 
\multicolumn{1}{|l|}{} & \ours\ ($t=16$)  & 97.55\% & 8944  & 0& 0.046  & 3.43  & 1.09    \\ \cline{2-8} 
\multicolumn{1}{|l|}{} & \ours\ ($t=103$)  & 97.37\% & 8911  & 0& 0.008   & 3.51  & 1.09   \\ \hline

\multicolumn{1}{|l|}{\multirow{6}{*}{HIV-5}}& Virus-VG & 96.85\% & 9632 & 2 & 0.332 &  1619.34 + 312.68 & \borja{26.83 | 0.64}  \\ \cline{2-8}
\multicolumn{1}{|l|}{} & VG-Flow & 96.87\% & 9625 & 2 & 0.331 & 1619.34 + 312.20 & \borja{26.83 | 0.65}\\ \cline{2-8} 
\multicolumn{1}{|l|}{} & \borjareview{PEHaplo} & 78.59\% & 9328  & 2 & 0.690   & 73.33 & 4.84 \\ \cline{2-8}
\multicolumn{1}{|l|}{} & \borjareview{PredictHaplo} & 99.90\% & 9663 & 0 & 0.591 & 4.00 + 120.13 & 1.05 \\\cline{2-8}
\multicolumn{1}{|l|}{} & \borjareview{CliqueSNV} & 99.86\% & 9649 & 0 & 1.15 & 4.00 + 93.67 & 8.51  \\\cline{2-8}
\multicolumn{1}{|l|}{} & viaDBG ($t=21$) & 97.50\%& 8046& 2& 0.151 &  62.34  & 2.89   \\ \cline{2-8} 
\multicolumn{1}{|l|}{} & viaDBG ($t=146$) & 95.27\% & 6237  & 3& 0.161  & 61.23 & 2.87   \\ \cline{2-8} 
\multicolumn{1}{|l|}{} & \ours\ ($t=21$)  & 95.58\% & 9617  & 2& 0.222 &  3.32  & 1.07   \\ \cline{2-8} 
\multicolumn{1}{|l|}{} & \ours\ ($t=146$)  & 99.71\% & 9237 & 2& 0.321  & 3.26 & 1.07   \\ \hline

\multicolumn{1}{|l|}{\multirow{6}{*}{POLIO-6}}& Virus-VG & 89.96\% & 7436 & 0 & 0.141 & 3455.00 + 201.23 & \borja{17.30 | 0.73}  \\ \cline{2-8}
\multicolumn{1}{|l|}{} & VG-Flow & 99.49\% & 7388 & 2 & 0.137 & 3455 + 532.33 & \borja{17.30 | 0.30} \\ \cline{2-8} 
\multicolumn{1}{|l|}{} & \borjareview{PEHaplo} & 98.15\% & 7428 & 0 & 0.125 & 107.96 & 3.63  \\\cline{2-8}
\multicolumn{1}{|l|}{} & \borjareview{PredictHaplo} & 49.81\% & 7428 & 0 & 0.646 & 82.35 & 0.92 \\\cline{2-8}
\multicolumn{1}{|l|}{} & \borjareview{CliqueSNV} & 83.07\% & 7428  & 0 & 1.84 & 27.95 & 8.45\\\cline{2-8}
\multicolumn{1}{|l|}{} & viaDBG ($t=12$) & 73.81\%& 1760 & 0 &  0.018  & 49.21  & 2.52   \\ \cline{2-8} 
\multicolumn{1}{|l|}{} & viaDBG ($t=51$) & 80.20\% & 2290  & 0 & 0.016  & 47.90 & 2.52   \\ \cline{2-8} 
\multicolumn{1}{|l|}{} & \ours\ ($t=12$)  & 86.90\% & 4540  & 0 & 0.105  & 3.21  & 1.07   \\ \cline{2-8} 
\multicolumn{1}{|l|}{} & \ours\ ($t=51$)  & 97.40\% & 7428 & 0 & 0.247 & 2.61 & 1.06   \\ \hline

\multicolumn{1}{|l|}{\multirow{6}{*}{ZIKV-15}}  & Virus-VG & 99.56\% & 10212   & 0 & 0.077 & 706 + 407.51  & \borja{13.45 | 1.37}  \\ \cline{2-8}
\multicolumn{1}{|l|}{} & VG-Flow & 83.05\% & 10210 & 0 & 0.144 & 706.00 + 406.22 & \borja{13.45 | 0.62}\\ \cline{2-8} 
\multicolumn{1}{|l|}{} & \borjareview{PEHaplo} & 98.32\% & 10247 & 0 & 2.05 & 321.53 & 0.08 | 8.80 \\ \cline{2-8}
\multicolumn{1}{|l|}{} & \borjareview{PredictHaplo} & 46.65\% & 10251 & 0 & 0.133  & 4.06 + 149.68 & 1.11\\\cline{2-8}
\multicolumn{1}{|l|}{} & \borjareview{CliqueSNV} & 66.66\% & 10251 & 0 & 0.036 & 4.06 + 126.28 & 8.38 \\\cline{2-8}
\multicolumn{1}{|l|}{} &  viaDBG ($t=5$) & 89.85\% & 1398 & 0& 0.110 & 65.48 & 3.25   \\ \cline{2-8} 
\multicolumn{1}{|l|}{} &  viaDBG ($t=26$) & 92.61\% & 2107  & 0 & 0.109 & 66.03 & 3.25   \\ \cline{2-8} 
\multicolumn{1}{|l|}{} &  \ours\ ($t=5$) &  80.92\% & 3042  & 0& 0.111 & 4.80  & 1.12   \\ \cline{2-8} 
\multicolumn{1}{|l|}{} & \ours\ ($t=26$)  & 99.08\% & 10140 & 0& 0.142 & 4.05 & 1.11   \\ \hline

\multicolumn{1}{|l|}{\multirow{6}{*}{HIV-real}} & Virus-VG & 83.36\% & 8637 & 0 & 3.384 & 3550.00 + 440.71 & \borja{26.85 | 0.80} \\ \cline{2-8}
\multicolumn{1}{|l|}{} & VG-Flow & 89.99\% & 5950 & 0 & 1.100 & 3550.00 + 1499.61 & \borja{26.85 | 1.47}\\ \cline{2-8}
\multicolumn{1}{|l|}{} & \borjareview{PEHaplo} & 91.43\% & 1262 & 0 & 0.074 & 68.34 & 3.48 \\\cline{2-8}
\multicolumn{1}{|l|}{} & \borjareview{PredictHaplo} & 90.21\% & 8702 & 0 & 0.287 & 4.71 + 100.75 & 0.87 \\\cline{2-8}
\multicolumn{1}{|l|}{} & \borjareview{CliqueSNV} & 72.17\% & 8676  & 0 & 1.125 & 4.71 + 136.68 & 9.03 \\\cline{2-8}
\multicolumn{1}{|l|}{} & viaDBG ($t=128$) & 87.25\% & 1813& 0& 0.197 & 17.24  & 3.74   \\ \cline{2-8} 
\multicolumn{1}{|l|}{} & viaDBG ($t=172$) & 89.36\% & 1670  & 0& 0.215& 17.24 & 3.75   \\ \cline{2-8} 
\multicolumn{1}{|l|}{} & \ours\ ($t=128$) & 90.27\% & 2302  & 1 & 0.349 & 3.78 & 1.07   \\ \cline{2-8} 
\multicolumn{1}{|l|}{} & \ours\ ($t=172$)  & 90.85\% & 2267 & 0& 0.292 & 3.73 & 1.07   \\ \hline
\end{tabular}
\end{table*}

\subsection{Haplotype abundance estimation}
In the experimental evaluation of the haplotype relative frequency estimation error, we use the following two measures:
\begin{itemize}
    \item Mean Estimation Error - which is the average error per haplotype  $$MEE = \frac{\sum_{c \in C} |\hat{freq_c} - freq_c|}{\#C},$$
    where $\hat{freq_c}$ is the estimated relative abundance for the haplotype $c$ and $freq_c$ is the known haplotype relative abundance, finally $\#C$ is the number of haplotypes under evaluation.
    \item \borja{Estimation Error Standard Quasideviation - which measures the amount of dispersion in the frequency errors estimation} $$\hat{S}_{EE} = \sqrt{\frac{\sum_{c\in C} (\epsilon_c - MEE)^2}{\#C-1}},$$ where $\epsilon_c$ is the error in the frequency estimated for the haplotype $c$.
\end{itemize}

\leena{Next, we show the full results for frequency estimation.}
\borja{Tables \ref{table:distrib_err_5}--\ref{table:distrib_err_6} show the frequency estimations and estimation errors of each haplotype of the different datasets obtained by each tool. 
Similar to the average results,
these tables show that both Virus-VG and VG-Flow have a slightly better performance on frequency estimation than ViQUF. The reason for this difference is more likely to be related with the raw data than the \leena{frequency estimation}
step. While VG-Flow and Virus-VG use preassembled contigs as input and the number of reads mapping to a contig as abundance estimation, \ours\ uses unitigs, which are faster to compute but shorter than contigs, and the average of the $k$-mer counting in each unitig as abundance estimation, once again faster to compute but in this case more error prone than mapping the reads. Therefore, \leena{because of the input data, the estimation of frequencies by \ours\ }are less reliable than the estimation by VG-Flow and Virus-Vg.}
\borja{According to our results, all the tools have good performance. \susana{There is a remarkable exception in the case of HIV-5 dataset, with a significantly high error for one of its haplotypes, thus, impacting the overall result.} Notice that the three methods make the same mistake overestimating the real frequency of the same haplotype, \textit{NL\_43}. This is probably because the \textit{NL\_43} haplotype shares most of its information with the rest of the haplotypes.
Apart from that, it is important to compare VG-Flow results on ZIKV-15 haplotype by haplotype because according to Table \ref{table:distrib_err_15} \leena{there are large differences in the error of frequency estimation from one haplotype to another.}
}
\begin{table*}[]
\centering
\resizebox{\textwidth}{!}{
\begin{tabular}{|l|r|r|r|r|r|r|r|r|r|r|r|}
\hline
Genome &  \multicolumn{1}{c|}{\begin{tabular}[c]{@{}c@{}}Real\\  Frequency\end{tabular}} & \multicolumn{1}{c|}{\begin{tabular}[c]{@{}c@{}}ViQUF \\ estimation\end{tabular}} & \multicolumn{1}{c|}{\begin{tabular}[c]{@{}c@{}}ViQUF\\  error\end{tabular}} & \multicolumn{1}{c|}{\begin{tabular}[c]{@{}c@{}}VG-Flow \\ estimation\end{tabular}} & \multicolumn{1}{c|}{\begin{tabular}[c]{@{}c@{}}VG-Flow \\ error\end{tabular}} & \multicolumn{1}{c|}{\begin{tabular}[c]{@{}c@{}}Virus-VG \\ estimation\end{tabular}} & \multicolumn{1}{c|}{\begin{tabular}[c]{@{}c@{}}Virus-VG \\ error\end{tabular}} &  \multicolumn{1}{c|}{\begin{tabular}[c]{@{}c@{}}PredictHaplo \\ estimation\end{tabular}} & \multicolumn{1}{c|}{\begin{tabular}[c]{@{}c@{}}PredictHaplo \\ error\end{tabular}} & \multicolumn{1}{c|}{\begin{tabular}[c]{@{}c@{}}CliqueSNV \\ estimation\end{tabular}} & \multicolumn{1}{c|}{\begin{tabular}[c]{@{}c@{}}CliqueSNV \\ error\end{tabular}} \\ \hline
5\_strain\_HIV\_NL43                                  & 11.20          & 15.71            & 4.51       & 24.24              & 13.04         & 26.66               & 15.46          & 27.00                                        & 15.80                                   & 12.97                                     & 1.77                                 \\ \hline
5\_strain\_HIV\_JRCSF                                 & 28.00          & 30.47            & 2.47        & 24.24              & 3.76          & 23.47               & 4.53           & 24.00                                        & 4.00                                    & 22.99                                     & 5.01                                 \\ \hline
5\_strain\_HIV\_YU2                                   & 11.10          & 5.56             & 5.53       & 5.24               & 5.86          & 5.08                & 6.02           & 5.00                                         & 6.10                                    & 4.90                                      & 6.20                                 \\ \hline
5\_strain\_HIV\_89.6                                  & 22.10          & 20.04            & 2.05        & 18.74              & 3.36          & 18.13               & 3.97           & 18.00                                        & 4.10                                    & 10.39                                     & 11.71                                \\ \hline
5\_strain\_HIV\_HXB2                                  & 27.30          & 28.20            & 0.91        & 27.54              & 0.24          & 26.66               & 0.64           & 26.00                                        & 1.30                                    & 16.86                                     & 10.44                                \\ \hline
\end{tabular}}
\caption{Frequency estimations and estimation errors for the HIV-5 dataset.}
\label{table:distrib_err_5}
\end{table*}

\begin{table*}[]
\centering
\resizebox{\textwidth}{!}{
\begin{tabular}{|l|r|r|r|r|r|r|r|r|r|r|r|}
\hline
Genome & \multicolumn{1}{c|}{\begin{tabular}[c]{@{}c@{}}Real\\  Frequency\end{tabular}} & \multicolumn{1}{c|}{\begin{tabular}[c]{@{}c@{}}ViQUF \\ estimation\end{tabular}} & \multicolumn{1}{c|}{\begin{tabular}[c]{@{}c@{}}ViQUF\\  error\end{tabular}} & \multicolumn{1}{c|}{\begin{tabular}[c]{@{}c@{}}VG-Flow \\ estimation\end{tabular}} & \multicolumn{1}{c|}{\begin{tabular}[c]{@{}c@{}}VG-Flow \\ error\end{tabular}} & \multicolumn{1}{c|}{\begin{tabular}[c]{@{}c@{}}Virus-VG \\ estimation\end{tabular}} & \multicolumn{1}{c|}{\begin{tabular}[c]{@{}c@{}}Virus-VG \\ error\end{tabular}} &  \multicolumn{1}{c|}{\begin{tabular}[c]{@{}c@{}}PredictHaplo \\ estimation\end{tabular}} & \multicolumn{1}{c|}{\begin{tabular}[c]{@{}c@{}}PredictHaplo \\ error\end{tabular}} & \multicolumn{1}{c|}{\begin{tabular}[c]{@{}c@{}}CliqueSNV \\ estimation\end{tabular}} & \multicolumn{1}{c|}{\begin{tabular}[c]{@{}c@{}}CliqueSNV \\ error\end{tabular}}\\ \hline
HCV\_EU155339.2                                       & 12.00          & 12.06            & 0.06        & 12.11              & 0.11          & 12.10               & 0.10           & 13.10                   & 1.10               & 0.00                    & 12.00               \\ \hline
HCV\_EU255981.1                                       & 13.00          & 12.69            & 0.31        & 12.96              & 0.04          & 12.96               & 0.04           & 5.00                    & 8.00               &  0.00                    & 13.00               \\ \hline
HCV\_EU255973.1                                       & 10.00          & 10.41            & 0.41        & 9.98               & 0.02          & 9.98                & 0.02           & 12.20                   & 2.20               & 0.00                    & 10.00               \\ \hline
HCV\_EU255980.1                                       & 5.00           & 5.13             & 0.13        & 5.03               & 0.03          & 5.03                & 0.03           & 6.10                    & 1.10               & 0.00                    & 5.00               \\ \hline
HCV\_EU155344.2                                       & 5.00           & 5.01             & 0.01        & 5.05               & 0.05          & 5.05                & 0.05           & 27.80                   & 22.80              & 0.00                    & 5.00               \\ \hline
HCV\_EU255989.1                                       & 19.00          & 18.98            & 0.02        & 18.89              & 0.11          & 18.88               & 0.12           & 10.10                   & 8.90               & 0.00                    & 19.00               \\ \hline
HCV\_EU255983.1                                       & 6.00           & 6.10             & 0.10        & 6.02               & 0.02          & 6.02                & 0.02           & 12.40                   & 6.40               & 0.00                    & 6.00               \\ \hline
HCV\_EU234065.2                                       & 8.00           & 7.97             & 0.03        & 8.00               & 0.00          & 8.00                & 0.00           & 0.00                    & 8.00               & 0.00                    & 8.00               \\ \hline
HCV\_EU255982.1                                       & 10.00          & 9.79             & 0.21        & 10.02              & 0.02          & 10.02               & 0.02           & 8.10                    & 1.90               & 0.00                    & 10.00               \\ \hline
HCV\_EU255965.1                                       & 12.00          & 11.81            & 0.19        & 11.95              & 0.05          & 11.95               & 0.05           & 5.00                    & 7.00               & 100.00               & 87.00           \\ \hline
\end{tabular}}
\caption{Frequency estimations and estimation errors for the HCV-10 dataset.}
\label{table:distrib_err_10}
\end{table*}

\begin{table*}[]
\centering
\resizebox{\textwidth}{!}{
\begin{tabular}{|l|r|r|r|r|r|r|r|r|r|r|r|}
\hline
 Genome & \multicolumn{1}{c|}{\begin{tabular}[c]{@{}c@{}}Real\\  Frequency\end{tabular}} & \multicolumn{1}{c|}{\begin{tabular}[c]{@{}c@{}}ViQUF \\ estimation\end{tabular}} & \multicolumn{1}{c|}{\begin{tabular}[c]{@{}c@{}}ViQUF\\  error\end{tabular}} & \multicolumn{1}{c|}{\begin{tabular}[c]{@{}c@{}}VG-Flow \\ estimation\end{tabular}} & \multicolumn{1}{c|}{\begin{tabular}[c]{@{}c@{}}VG-Flow \\ error\end{tabular}} & \multicolumn{1}{c|}{\begin{tabular}[c]{@{}c@{}}Virus-VG \\ estimation\end{tabular}} & \multicolumn{1}{c|}{\begin{tabular}[c]{@{}c@{}}Virus-VG \\ error\end{tabular}} &  \multicolumn{1}{c|}{\begin{tabular}[c]{@{}c@{}}PredictHaplo \\ estimation\end{tabular}} & \multicolumn{1}{c|}{\begin{tabular}[c]{@{}c@{}}PredictHaplo \\ error\end{tabular}} & \multicolumn{1}{c|}{\begin{tabular}[c]{@{}c@{}}CliqueSNV \\ estimation\end{tabular}} & \multicolumn{1}{c|}{\begin{tabular}[c]{@{}c@{}}CliqueSNV \\ error\end{tabular}} \\ \hline
Strain 0                                              & 2.00           & 2.01             & 0.01        & 0.00               & 2.00          & 2.05                & 0.05           & 12.40                   & 10.40              & 0.00                 & 2.00            \\ \hline
Strain 1                                              & 2.00           & 2.01             & 0.01        & 2.10               & 0.10          & 2.05                & 0.05           & 0.00                    & 2.00               & 0.00                 & 2.00            \\ \hline
Strain 2                                              & 2.00           & 2.01             & 0.01        & 8.31               & 6.31          & 1.98                & 0.02           & 32.40                   & 30.40              & 0.00                 & 2.00            \\ \hline
Strain 3                                              & 4.00           & 4.03             & 0.03        & 2.19               & 1.81          & 2.30                & 1.70           & 0.00                    & 4.00               & 0.00                 & 4.00            \\ \hline
Strain 4                                              & 4.00           & 4.27             & 0.27        & 4.42               & 0.42          & 4.35                & 0.35           & 0.00                    & 4.00               & 0.00                 & 4.00            \\ \hline
Strain 5                                              & 4.00           & 4.09             & 0.09        & 4.83               & 0.83          & 4.29                & 0.29           & 0.00                    & 4.00               & 3.20                 & 0.80            \\ \hline
Strain 6                                              & 6.00           & 6.02             & 0.02        & 2.77               & 3.23          & 4.87                & 1.13           & 0.00                    & 6.00               & 6.50                 & 0.50            \\ \hline
Strain 7                                              & 6.00           & 6.21             & 0.21        & 4.54               & 1.46          & 4.58                & 1.42           & 12.10                   & 6.10               & 6.60                 & 0.60            \\ \hline
Strain 8                                              & 6.00           & 6.26             & 0.26        & 0.00               & 6.00          & 3.66                & 2.34           & 0.00                    & 6.00               & 6.50                 & 0.50            \\ \hline
Strain 9                                              & 8.00           & 8.22             & 0.22        & 8.88               & 0.88          & 8.76                & 0.76           & 8.10                    & 0.10               & 8.10                 & 0.10            \\ \hline
Strain 10                                             & 8.00           & 8.39             & 0.39        & 8.95               & 0.95          & 8.84                & 0.84           & 8.00                    & 0.00               & 8.00                 & 0.00            \\ \hline
Strain 11                                             & 8.00           & 8.00             & 0.00        & 8.93               & 0.93          & 8.77                & 0.77           & 0.00                    & 8.00               & 7.90                 & 0.10            \\ \hline
Strain 12                                             & 13.00          & 13.81            & 0.81        & 14.68              & 1.68          & 14.54               & 1.54           & 13.60                   & 0.60               & 13.38                & 0.38            \\ \hline
Strain 13                                             & 13.00          & 13.70            & 0.70        & 14.72              & 1.72          & 14.47               & 1.47           & 13.40                   & 0.40               & 13.36                & 0.36            \\ \hline
Strain 14                                             & 13.00          & 13.79            & 0.79        & 14.69              & 1.69          & 14.49               & 1.49           & 0.00                    & 13.00              & 13.36                & 0.36            \\ \hline
\end{tabular}}
\caption{Frequency estimations and estimation errors for the ZIKV-15 dataset.}
\label{table:distrib_err_15}
\end{table*}

\begin{table*}[]
\centering
\resizebox{\textwidth}{!}{
\begin{tabular}{|l|r|r|r|r|r|r|r|r|r|r|r|}
\hline
Genome & \multicolumn{1}{c|}{\begin{tabular}[c]{@{}c@{}}Real\\  Frequency\end{tabular}} & \multicolumn{1}{c|}{\begin{tabular}[c]{@{}c@{}}ViQUF \\ estimation\end{tabular}} & \multicolumn{1}{c|}{\begin{tabular}[c]{@{}c@{}}ViQUF\\  error\end{tabular}} & \multicolumn{1}{c|}{\begin{tabular}[c]{@{}c@{}}VG-Flow \\ estimation\end{tabular}} & \multicolumn{1}{c|}{\begin{tabular}[c]{@{}c@{}}VG-Flow \\ error\end{tabular}} & \multicolumn{1}{c|}{\begin{tabular}[c]{@{}c@{}}Virus-VG \\ estimation\end{tabular}} & \multicolumn{1}{c|}{\begin{tabular}[c]{@{}c@{}}Virus-VG \\ error\end{tabular}} &  \multicolumn{1}{c|}{\begin{tabular}[c]{@{}c@{}}PredictHaplo \\ estimation\end{tabular}} & \multicolumn{1}{c|}{\begin{tabular}[c]{@{}c@{}}PredictHaplo \\ error\end{tabular}} & \multicolumn{1}{c|}{\begin{tabular}[c]{@{}c@{}}CliqueSNV \\ estimation\end{tabular}} & \multicolumn{1}{c|}{\begin{tabular}[c]{@{}c@{}}CliqueSNV \\ error\end{tabular}} \\ \hline
seq\_1                                                & 50.80          & 46.98            & 3.81        & 50.86              & 0.06          & 49.32               & 1.48           & 77.20                   & 26.40              & 54.26                & 3.46            \\ \hline
seq\_2                                                & 25.40          & 21.81            & 3.58        & 18.84              & 6.56          & 22.55               & 2.85           & 0.00                    & 25.40              & 10.00                & 15.40           \\ \hline
seq\_3                                                & 12.70          & 10.97            & 1.72        & 15.91              & 3.21          & 15.33               & 2.63           & 14.00                   & 1.30               & 9.62                 & 3.08            \\ \hline
seq\_4                                                & 6.30           & 9.00             & 2.70        & 8.66               & 2.36          & 8.35                & 2.05           & 0.00                    & 6.30               & 4.38                 & 1.92            \\ \hline
seq\_5                                                & 3.20           & 8.99             & 5.79        & 3.45               & 0.25          & 3.21                & 0.01           & 8.70                    & 5.50               & 6.20                 & 3.00            \\ \hline
seq\_6                                                & 1.60           & 2.23             & 0.69        & 2.27               & 0.67          & 1.24                & 0.36           & 0.00                    & 1.60               & 0.00                 & 1.60            \\ \hline
\end{tabular}}
\caption{Frequency estimations and estimation errors for the POLIO-6 dataset.}
\label{table:distrib_err_6}
\end{table*}

\subsection{Comparison using precision and recall}

\borjareview{In the main paper, the metrics we employed are generalizations of common assembly quality metrics used for \textit{de novo} genome assemblers. Recently,  new approaches have been suggested to give a new point of view on the quality of the assembly. 
 Table \ref{table:ref_metrics}  shows the results of   an adaption of the well-known computational metrics  \textit{precision} and \textit{recall}, suggested by \cite{PMID:32151775}. }

\borjareview{Before explaining how the metrics are adapted, we have to introduce the term \textit{proper contigs}. The set of proper contigs $Q'$ is a subset of the original set of contigs $Q$, namely $Q'\subseteq Q$, where $Q' = \{\forall q \in Q/ mismatches(q) \leq 1\%\}$. Then, precision is defined as $\frac{TP}{TP+FP}$, where $TP$ (true positives) is the total frequency of the haplotypes correctly predicted by the set of proper contigs  and $FP$ (false positives) is the total frequency of contigs $q \in Q'$ which do not match with any haplotype. On the other hand, recall is $\frac{TP}{TP+FN}$ where $FN$ (false negatives) are $1-TP$, thus $recall = TP$.}

\borjareview{Since the results in Table \ref{table:ref_metrics} come from the same contigs than the ones in the main paper, similarities are expected. However, we can see new insights that might be interesting to point out. First, Virus-VG and VG-Flow exhibit a slightly better behavior since the haplotypes that they lose are the least abundant; thus, precision and recall suffer less than the metrics in the main paper, where all haplotypes count the same. Furthermore, the precision and recall of ViQUF and PEHaplo for HIV-5 dataset are worse since both retrieve some contigs with high level of mismatches  (1\% to 2.7\%), and these contigs align best against 5\_strain\_HIV\_JRCSF, which is a high abundance haplotype. The results for reference-based method are stable and have no remarkable changes from the previously exposed.}


\josereview{
These metrics are more geared towards reference-based methods, albeit \cite{PMID:32151775} extend them to \textit{de novo} constructions.
Among other problems, they fail to measure how fragmented the assemblies are, as this is not an issue for reference-based methods, which always retrieve full haplotypes.
They prioritize the contigs of the most abundant haplotypes, a characteristic inherit from their original definition,  but in viral quasispecies reconstruction, it is very important to also retrieve the low abundance haplotypes correctly.
However, we included them to give a broader picture of our method.
}

\begin{table*}[]
\centering
\begin{tabular}{l|r|r|r|r|r|r|r|l|}
\cline{2-9}
                                   & \multicolumn{4}{c|}{Precision}                                                                                        & \multicolumn{4}{c|}{Recall}                                                                       \\ \cline{2-9} 
                                   & \multicolumn{1}{l|}{HIV-5} & \multicolumn{1}{l|}{HCV-10} & \multicolumn{1}{l|}{POLIO-6} & \multicolumn{1}{l|}{ZIKV-15} & \multicolumn{1}{l|}{HIV-5} & \multicolumn{1}{l|}{HCV-10} & \multicolumn{1}{l|}{POLIO-6} & ZIKV-15 \\ \hline
\multicolumn{1}{|l|}{Virus-VG}     & 100.00\%                   & 100.00\%                    & 100.00\%                     & 100.00\%                     & 95.68\%                    & 99.79\%                     & 98.31\%                      & 98.62\% \\ \hline
\multicolumn{1}{|l|}{VG-Flow}      & 100.00\%                   & 100.00\%                    & 100.00\%                     & 100.00\%                     & 95.60\%                    & 99.81\%                     & 99.70\%                      & 88.82\% \\ \hline
\multicolumn{1}{|l|}{PEHaplo}      & 87.18\%                    & 100.00\%                    & 100.00\%                     & 70.50\%                      & 75.46\%                    & 99.66\%                     & 99.42\%                      & 97.60\% \\ \hline
\multicolumn{1}{|l|}{PredictHaplo} & 49.26\%                    & 100.00\%                    & 85.71\%                      & 100.00\%                     & 48.16\%                    & 91.99\%                     & 76.20\%                      & 55.99\% \\ \hline
\multicolumn{1}{|l|}{CliqueSNV}    & 55.20\%                    & 0.00\%                      & 51.29\%                      & 100.00\%                     & 88.47\%                    & 0.00\%                      & 63.20\%                      & 84.99\% \\ \hline
\multicolumn{1}{|l|}{ViQUF}        & 71.99\%                    & 100.00\%                    & 100.00\%                     & 94.26\%                      & 70.51\%                    & 98.53\%                     & 99.69\%                      & 97.95\% \\ \hline
\end{tabular}
\caption{Precision and recall measures for the viral quasispecies tools, de novo and reference-based}
\label{table:ref_metrics}
\end{table*}
\bibliographystyle{IEEEtran}
\bibliography{bibliography.bib}